\documentclass[a4paper,nofootinbib,useAMS,usenatbib]{mn2e}
\usepackage{amssymb}
\usepackage{amsmath}
\usepackage{epsfig}
\usepackage{subfigure}
\usepackage{mathrsfs}
\usepackage{longtable}
\usepackage[usenames,dvipsnames]{xcolor}
\usepackage{mathtools}
\usepackage{relsize}
\usepackage{geometry}
\usepackage{graphicx}

\begin{document}
%%%%%%%%%%%%% colortext comments %%%%%%%%%%%%%%
\newcommand{\tkDM}[1]{\textcolor{red}{#1}}                     % Doddy
\newcommand{\unit}[1]{\ensuremath{\, \mathrm{#1}}}
\newcommand{\apj}{Astrophys. J.}
\newcommand{\mnras}{Mon. Not. R. Astron. Soc.}

%%%%%%%%%%%%%%%%%%%%%%%%%%%%%%%%%%%%%%%%%%

\title{A Model For Halo Formation With Axion Mixed Dark Matter}

\author[D. J. E. Marsh and J. Silk] {David~J.~E.~Marsh,$^1$\thanks{dmarsh@perimeterinstitute.ca} Joseph~Silk.$^{2,3,4}$\\
$^1$Perimeter Institute, 31 Caroline St N,  Waterloo, ON, N2L 6B9, Canada\\ 
$^2$Institut dÕAstrophysique, UMR 7095 CNRS, Universit\'{e} Pierre et Marie Curie, 98bis Blvd Arago, 75014 Paris, France \\
$^3$Department of Physics and Astronomy, The Johns Hopkins University, Homewood Campus, Baltimore MD 21218, USA \\
$^4$Beecroft Institute of Particle Astrophysics and Cosmology, Department of Physics, University of Oxford, Oxford OX1 3RH, UK
}
\date{\today}
\maketitle

%---------------------- ABSTRACT -------------------------
\begin{abstract}
There are several issues to do with dwarf galaxy predictions in the standard $\Lambda$CDM cosmology that have suscitated much recent debate about the possible modification of the nature of dark matter as providing a solution. We explore a novel solution involving ultra-light axions that can potentially resolve the missing satellites problem, the cusp-core problem, and the `too big to fail'  problem. We discuss approximations to non-linear structure formation in dark matter models containing a component of ultra-light axions across four orders of magnitude in mass, $10^{-24}\text{ eV}\lesssim m_a \lesssim 10^{-20}\text{ eV}$, a range too heavy to be well constrained by linear cosmological probes such as the CMB and matter power spectrum, and too light/non-interacting for other astrophysical or terrestrial axion searches. We find that an axion of mass $m_a\approx 10^{-21}\text{ eV}$ contributing approximately 85\% of the total dark matter can introduce a significant kpc scale core in a typical Milky Way satellite galaxy in sharp contrast to a thermal relic with a transfer function cut off at the same scale, while still allowing such galaxies to form in significant number. Therefore ultra-light axions do not suffer from the \emph{Catch 22} that applies to using a warm dark matter as a solution to the small scale problems of cold dark matter. Our model simultaneously allows formation of enough high redshift galaxies to allow reconciliation with observational constraints, and also reduces the maximum circular velocities of massive dwarfs so that baryonic feedback may more plausibly resolve the predicted overproduction of massive MWG dwarf satellites.
\end{abstract}


\section{Introduction}
\label{intro}

There are three outstanding problems in the dwarf galaxy astrophysics of the standard $\Lambda$CDM cosmology. The controversies on small scales may be summarised as a) the  Missing Satellites problem (MSP), b) the Cusp-Core problem (CCP), and c) the `too big to fail'  problem (which we refer to here as the Massive Failures Problem, MFP), all reviewed in \cite{2013arXiv1306.0913W}.

The MSP and CCP with CDM structure formation can both be solved by introducing a length scale into the DM. This can be thermal, coming from free-streaming of warm dark matter (WDM), or, as we will discuss below, non-thermal, coming from coherent oscillations of a light scalar field. The thermal solution may suffer from a \emph{Catch 22} issue, whereby galaxy formation occurs too late \citep{2012MNRAS.424.1105M}.
We show here that  the non-thermal solution both avoids this dilemma and also augurs well for a particle-orientated solution of MFP, a problem for which a feedback solution seems questionable \citep{2013arXiv1301.3137G, 2012MNRAS.422.1203B}, although not all agree
\citep{2013ApJ...765...22B}. 

The paper is organised as follows. We introduce ultra-light scalar dark matter and compare the linear theory to WDM in Section~\ref{sec:lin_power}. In Section~\ref{sec:hmf} we compute the halo mass function and model a cut-off in it. In Section~\ref{sec:density_profile} we discuss the halo-density profile and core formation. In Section~\ref{sec:tbtf} we discuss the relevance of our model for MFP, and in Section~\ref{sec:collapse} we discuss implications for high-redshift galaxy formation. The casual reader can skip to Section~\ref{sec:discussion} where we summarise and discuss our main results, and provide a guide to the relevant figures. Appendix~\ref{appendix:normalisation} provides details of our two-component density profile model.

\section{Ultra-light Scalar Dark Matter}
\label{sec:lin_power}

A coherently oscillating scalar field, $\phi$, in a quadratic potential $V=m_a^2\phi^2/2$, has an energy density that scales as $a^{-3}$ and thus can behave in cosmology as DM \citep{turner1983b,turner1986}\footnote{Exponential potentials can also be relevant in the tracking solution \cite{ferreira1997,ferreira1998}. Light fields as DM with various potentials have had their background evolution studied in e.g.\citep{matos2009}.}. The relic density contains a non-thermal component produced by vacuum realignment, which depends on the initial field displacement $\phi_i$. The Klein-Gordon equation is:
\begin{equation}
\ddot{\phi}+3H\dot{\phi}+m_a^2\phi=0\, ,
\label{eqn:klein_gordon}
\end{equation}
where the Hubble rate $H=\dot{a}/a$. When $H\gg m_a$ the field is frozen by Hubble friction and behaves as a contribution to the cosmological constant (which is negligible for sub-Planckian field values). Therefore, in order to contribute to DM we must have $m_a\gtrsim H_0\sim 10^{-33}\unit{eV}$. Once the mass overcomes the Hubble friction at $m_a\approx 3H(a_{\text{osc}})$ the field begins to coherently oscillate. The relic density is then an environmental variable set by the initial field displacement: $\Omega_a=\Omega_a(m_a,\phi_i)\approx (8 \pi G/3 H_0^2) a_{\rm osc}(m_a)^3 m_a^2 \phi_i^2/2$.

As we will discuss below there is a length scale, which depends on the inverse mass, below which perturbations in the scalar field energy density will not cluster. Therefore the clustering of light scalar DM is observationally analogous to that of thermal relic dark matter. In the range $10^{-33}\text{ eV}\leq m_a \lesssim 10^{-28}\text{ eV}$ the clustering scale is analogous to hot (H)DM, for example composed of thermal relic neutrinos of mass $m_\nu\lesssim 1\text{ eV}$ \citep{amendola2005,marsh2011b}. In this section we will discuss how scalar masses in the range $10^{-24} \text{ eV}\lesssim m_a \lesssim 10^{-20}\text{ eV}$ lead to structure formation that is analogous to WDM in an observationally relevant mass range. Related aspects of structure formation for axion/scalar dark matter in this mass range have been studied in, e.g. \cite{hu2000,matos2000,arbey2001,arbey2003,bernal2006,lee2009,park2012}. 

While the signatures of thermal relics and ultra-light scalars are similar in large scale structure, there are distinct signatures in the adiabatic and isocurvature CMB spectra \citep{marsh2011b,marsh2013}, and future measurements of weak lensing tomography can further break degeneracies \citep{euclid_theory_report}. For the range of axion masses we consider $a_{\rm osc}\propto (m_a/\text{eV})^{-1/2}$ and the redshift $z_{\rm osc}$ is in the range $10^5 \lesssim z_{\rm osc}\lesssim 10^7$. If the axion field is coupled to photons, the rolling field from $\phi=\phi_i$ to $\phi=0$ at $z_{\rm osc}$ can further affect the CMB. For axions with mass $m_a\lesssim 10^{-28}\text{ eV}$ that roll after recombination this leads to rotation of CMB polarisation (see e.g. \cite{5yearWMAP}). For heavier axions rolling at $z\sim 10^6$, photon production in primordial magnetic fields may lead to CMB spectral distortions \citep{mirizzi2009}. 
It is necessary to understand the observational signatures of the parameters $m_a$ and $\Omega_a$ if we are to make inferences about the nature of the DM from cosmological constraints, and in particular if hints from the CMB and large scale structure are pointing to a hot, warm, or ultra-light scalar component to the DM. In the rest of this article we will explore in detail structure formation with ultra-light scalars, and similarities and differences with thermal DM.

Such ultra-light scalars might arise in a string theory context. It is well known that string theory compactified on sufficiently complicated six-dimensional manifolds contains many axion like particles (ALPs) \citep{witten1984,witten2006}. In \cite{axiverse2009} it was pointed out that since the masses of such axions depend exponentially on the areas of the cycles in the compact manifold, one should expect a uniform distribution of axion masses on a logarithmic scale spanning many orders of magnitude. This phenomenon was dubbed the ``\emph{String Axiverse}''. Explicit constructions of the axiverse have been made in M-theory \citep{acharya2010a} and Type IIB theory \citep{cicoli2012c}.

The string axiverse has the potential to provide an elegant solution to the MSP, CCP, and possibly MFP, by leading us to expect as natural an axion mixed DM (aMDM) model with many axionic components populating hierarchically different mass regimes. Since the relic density produced via vacuum misalignment is environmental it can be taken as a free parameter to be constrained observationally, although theoretical priors can be considered \citep[e.g. ][]{aguirre2004,tegmark2006}. A component of CDM in the aMDM model from the axiverse arises naturally in the form of the QCD axion \citep{pecceiquinn1977,weinberg1978,wilczek1978,wise1981,preskill1983,berezhiani1992}. The mass of the QCD axion is fixed by the pion mass and decay constant, and the axion decay constant $f_a$. For stringy values of $f_a\sim 10^{16}\text{GeV}$ the QCD axion has a mass around $10^{-10} \text{eV}$. This is light, but not so light that the sound speed (see below) plays a cosmological role. The requirement that the QCD axion remains light enough, barring accidents, coincidences or fine-tuning, to solve the strong \emph{CP} problem is what guarantees the lightness of the other axions, and as such one should always expect some CDM component alongside the ultra-light ALPs (ULAs). Axion mixed dark matter with a QCD axion and supersymmetric neutralino is also expected in many models of beyond the standard model particle physics \citep[see, e.g. ][]{bae2013}.

\subsection{Transfer Functions}

The ULA perturbations, $\delta \phi$, do not behave as CDM: they have a non-zero sound speed\footnote{This is in the cosmological frame, e.g. synchronous or Newtonian gauge, where $\delta \phi\neq 0$.} $c_a^2=\delta P/\delta \rho$, which is scale-dependent and given by \citep{hu2000,hwang2009,marsh2010}:
\begin{eqnarray}
c_a^{2}=\left\{\begin{array}{ll}
\frac{k^2}{4m_a^2a^2}&\mbox{if $k\ll 2m_{a}a$},\\
1&\mbox{if $k\gg 2m_{a}a$}.\end{array}\right.\label{heuristic_cs}
\end{eqnarray}

One finds that in a cosmology where axions make up a fraction of the DM, $f_{ax}=\Omega_a/\Omega_d$, that this causes suppression of the matter power spectrum relative to the case where the DM is pure CDM. Suppression occurs for those modes $k$ that entered the horizon when the sound speed was large. The suppression is centred around a scale $k_m$, which depends on the mass, and takes the power spectrum down to some value $S$, which depends on $\Omega_a$, times its value in the CDM case. %The scale $k_m$ is a manifestation of the axion Jeans scale, $k_J$, and in Section~\ref{sec:density_profile} we will use it to set the expected size of cores in axion dominated halos.
%Although it is possible \citep{amendola2005,marsh2010} to estimate the values of $k_m$, $S$ etc. and even the shape of the axion transfer function from the axion mass and initial misalignment angle, we will not use analytic fits in this work, since they are not valid over the full range of $\{m_a,\Omega_a\}$ parameter space.

We compute the transfer functions and matter power spectrum in cosmologies containing CDM plus a ULA component using a modified version of the publicly available Boltzmann code \texttt{CAMB} \citep{lewis2000,camb}. The modification, which makes use of the fluid treatment of axion perturbations, is described in \cite{marsh_inprep}. The transfer function is defined as
\begin{equation}
T_{\rm ax}(k)=\left ( \frac{P_{\rm aMDM} (k)}{P_{\Lambda {\rm CDM}}(k)} \right)^{0.5} \, .
\label{eqn:transfer_ax}
\end{equation}
We compare to the WDM transfer function (in the case where all the DM is warm)
\begin{equation}
T_{\rm WDM}(k)=(1+(\alpha k)^{2\mu})^{-5/ \mu} \, ,
\label{eqn:tk_wdm}
\end{equation}
where $\mu=1.12$ \citep{bode2001}. The mixed C+WDM case is discussed in more detail in e.g. \cite{anderhalden2012}. A well-defined characteristic scale to assign to any such step-like transfer function is the `half-mode'
\begin{equation}
T(k_m)=0.5 (1- T(k\to \infty)) \, ,
\label{eqn:km_def}
\end{equation}
where $T(k\to\infty)\geq 0$ is the constant plateau value of the transfer function on small scales. This is not the Jeans scale where all structure is suppressed. The Jeans scale is found analytically to be \citep{hu2000}
\begin{equation}
k_J= (16 \pi G \rho_m)^{1/4}m_a^{1/2} \, .
\label{eqn:hu_jeans_def}
\end{equation} 
The $\rho^{1/4}$ scaling follows from balancing the growing and oscillating modes in $e^{\Gamma t}$ where $\Gamma^2=4\pi G\rho-(k^2/2m)^2$, with $k^2/2m$ coming from the oscillation frequency of the free field.

\begin{figure}
\includegraphics[scale=0.35]{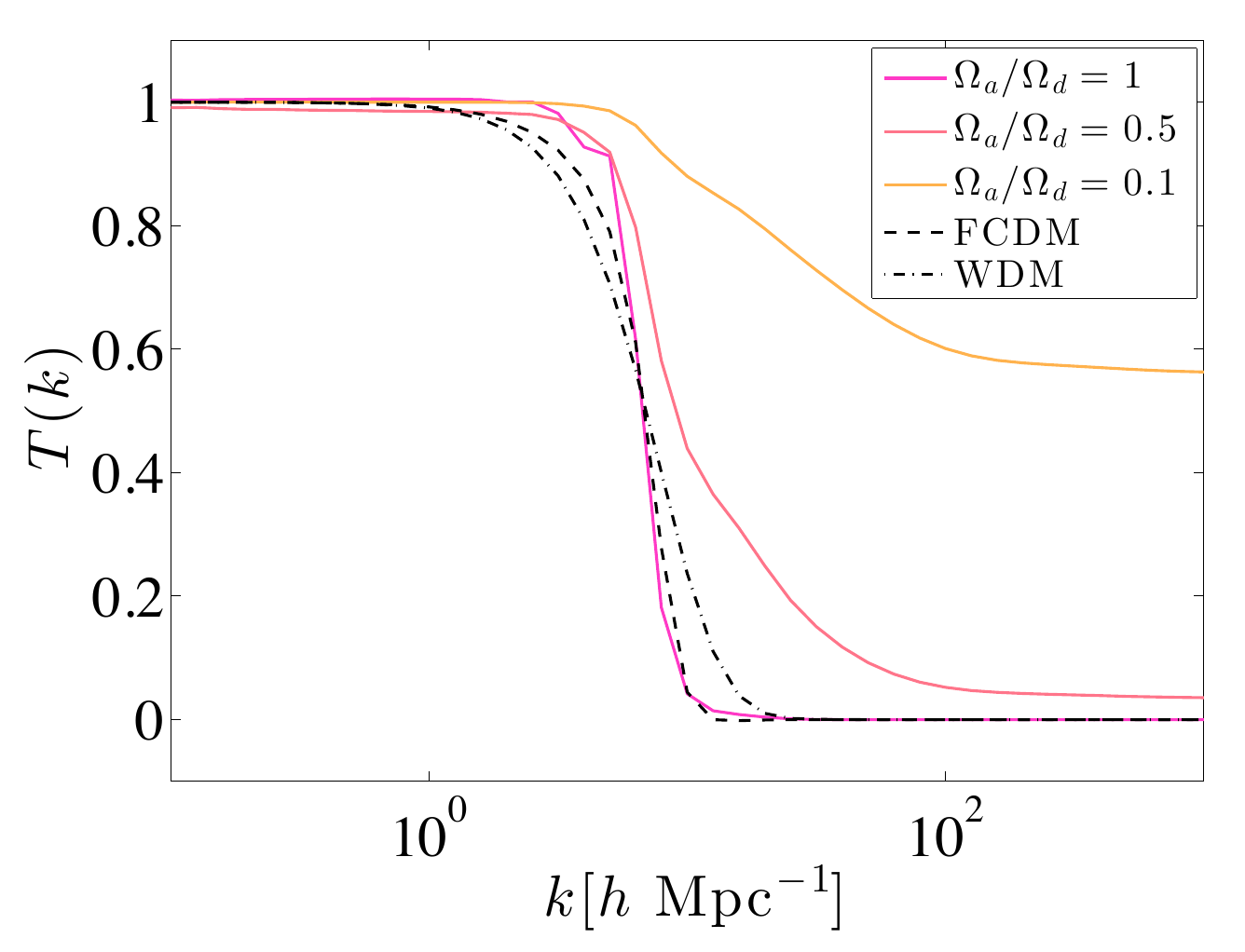}
\caption{The transfer function, Eq.~(\ref{eqn:transfer_ax}) for aMDM with $m_a=10^{-22}\text{ eV}$ and varying axion fractions to total DM. For comparison, we also plot the FCDM transfer function of \citet{hu2000} and the WDM transfer function (Eq.~(\ref{eqn:tk_wdm})) with $m_W\approx 0.84 \text{ keV}$ chosen to match the transfer function half mode, $k_m$ (Eq.~(\ref{eqn:km_def})). With this choice and $\Omega_a/\Omega_d=1$ the axion transfer function at $k_m$ is much steeper than its WDM counterpart.}\label{fig:lin_tk}
\end{figure}

In Fig.~\ref{fig:lin_tk} we plot the linear theory aMDM transfer function for a variety of aMDM models, all with $m_a=10^{-22}\text{ eV}$ which gives $k_m(10^{-22}\text{ eV})=6.7 \,h\text{ Mpc}^{-1}$. We compare to Eq.~(\ref{eqn:tk_wdm}) with $\alpha\approx 0.065\, h^{-1}\text{Mpc}$ chosen to give the same $k_m$. Taking $\Omega_d h^2=0.112$, $\Omega_b h^2=0.0226$, $h=0.7$ as our benchmark cosmology, \cite{angulo2013} gives $\alpha$ in terms of the WDM mass as:
\begin{equation}
\alpha=0.052 \left(\frac{m_W}{\text{keV}} \right)^{-1.15} \, h^{-1}\text{ Mpc}\, .
\end{equation}
Therefore, the matching of half-mode scales gives $m_W\approx 0.83 \text{ keV}$ as equivalent to $m_a=10^{-22}\text{ eV}$.

The logarithmic slope, $\mathrm{d}\ln T(k)/\mathrm{d}\ln k$, evaluated at $k=k_m$ is much steeper for the pure axion model than for the pure WDM model, in agreement with the transfer function of \cite{hu2000} for `Fuzzy' (F)CDM, also shown in Fig.~\ref{fig:lin_tk}. 

With decreasing fraction of DM in ULAs the slope becomes shallower, and $k_m$ moves out to larger values. The steeper slope for ULAs compared to WDM means that models with the same half-mode will not have $k_J=k_{FS}$ (where $k_{FS}$ is the WDM free-streaming scale, which some authors define differently), and vice versa. Matching Jeans and free-streaming scales, axions will have more power on larger scales relative to WDM; matching the half-mode, axions will have less power on small scales relative to WDM. We choose always to match the transfer function half-mode, since it is well-defined for both models.

\subsection{Mass Scales}
\label{sec:mass_scales}

We associate characteristic masses to scales $k$ through the mass enclosed within a sphere of radius the half wavelength $\lambda/2=\pi/k$:
\begin{equation}
M=\frac{4}{3}\pi \left( \frac{\lambda}{2} \right)^3 \rho_0 \, ,
\label{eqn:suppress_mass}
\end{equation}
where $\rho_0$ is the matter density. 

\begin{figure}
\includegraphics[scale=0.3]{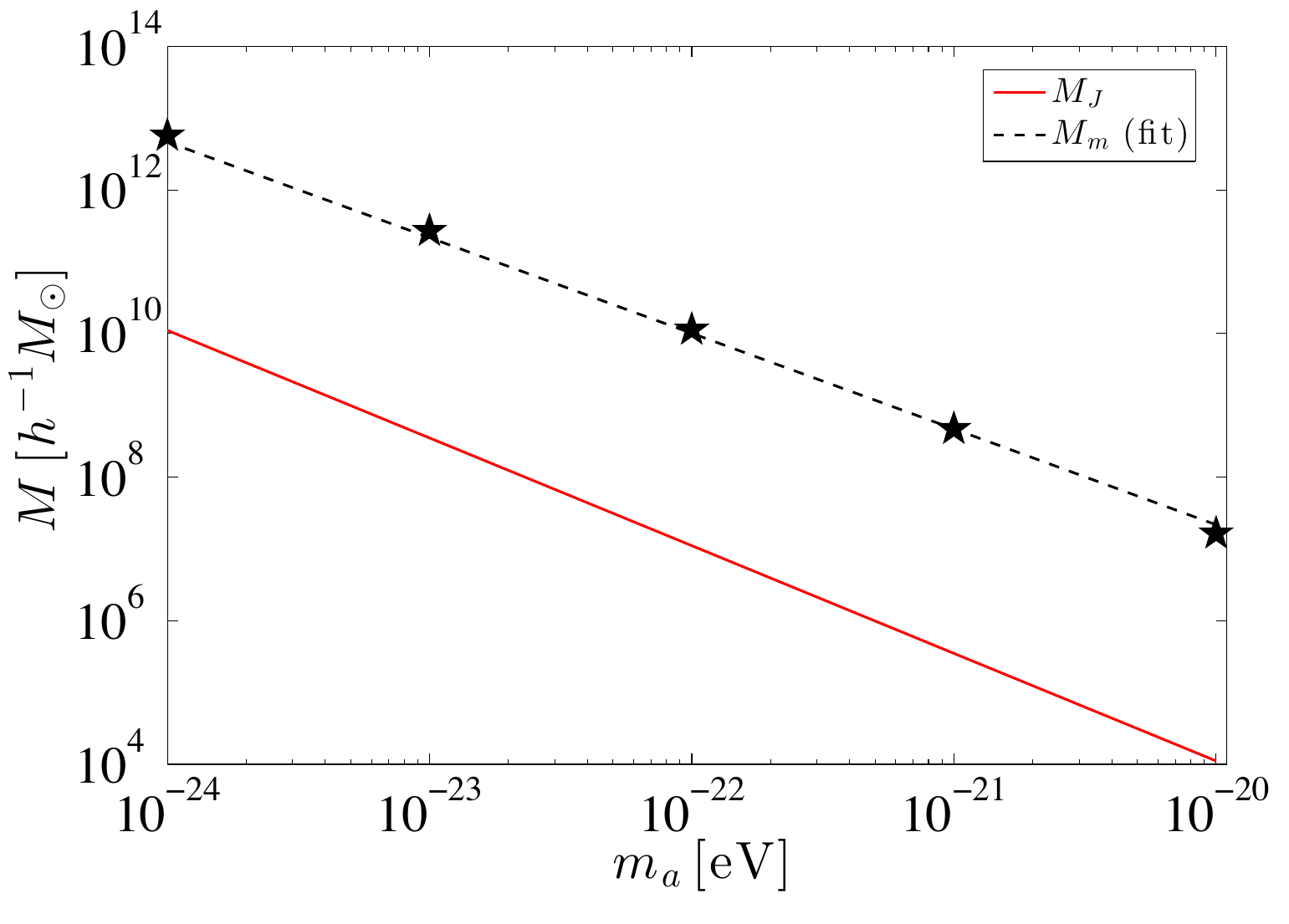}
\caption{The characteristic mass associated to the half-mode, $k_m$, of the transfer function as a function of axion mass, $M_m(m_a)$, found from Eqs.~(\ref{eqn:km_def}) and (\ref{eqn:suppress_mass}). It is well fit by a power law $M_m\propto m_a^{-\gamma}$ with $\gamma \approx 1.35$. We also show the mass associated to the Jeans scale of Eq.~(\ref{eqn:hu_jeans_def}), which is lower by two to three orders of magnitude.}\label{fig:characteristic_mass}
\end{figure}

In particular, using $k=k_m$ we can expect suppression of the formation of halos below $M_m$ caused by the decrease in linear power on these scales. We have shown the effects in the transfer function with low axion fraction for illustration, but as we will see in Section~\ref{sec:density_profile} the only axion fractions relevant for producing cored density profiles are large, $\Omega_a/\Omega_m\gtrsim 0.85$, and so the characteristic scales will be very close to their values for the pure ULA DM case. Mass scales relevant for halo formation cover axions in the range $10^{-24}\text{ eV}\lesssim m_a \lesssim 10^{-20}\text{eV}$. Axions lighter than this are well probed by the CMB and the linear matter power spectrum \citep{amendola2005,marsh_inprep}, while those heavier are probed by supermassive black holes \citep{arvanitaki2010,pani2012} and terrestrial experiments \citep{jaeckel2010b,ringwald2012,ringwald2012b}. In Fig.~\ref{fig:characteristic_mass} we plot $M_m(m_a)$ for the pure ULA cosmology and find it to be fit well by a power law $M_m\propto m_a^{-\gamma}$ with $\gamma\approx 1.35$ by least squares over the range of interest. This is very close to the value $\gamma=4/3$ using the fit of \cite{hu2000}. 

In Fig.~\ref{fig:characteristic_mass} we also show the Jeans mass, $M_J$, which is lower by more than two orders of magnitude than $M_m$. The axion Jeans scale is analogous to the WDM free-streaming scale, where $M_{\rm fs}$ is also some orders of magnitude lower than $M_m$ \citep{angulo2013}.

\begin{figure}
\includegraphics[scale=0.35]{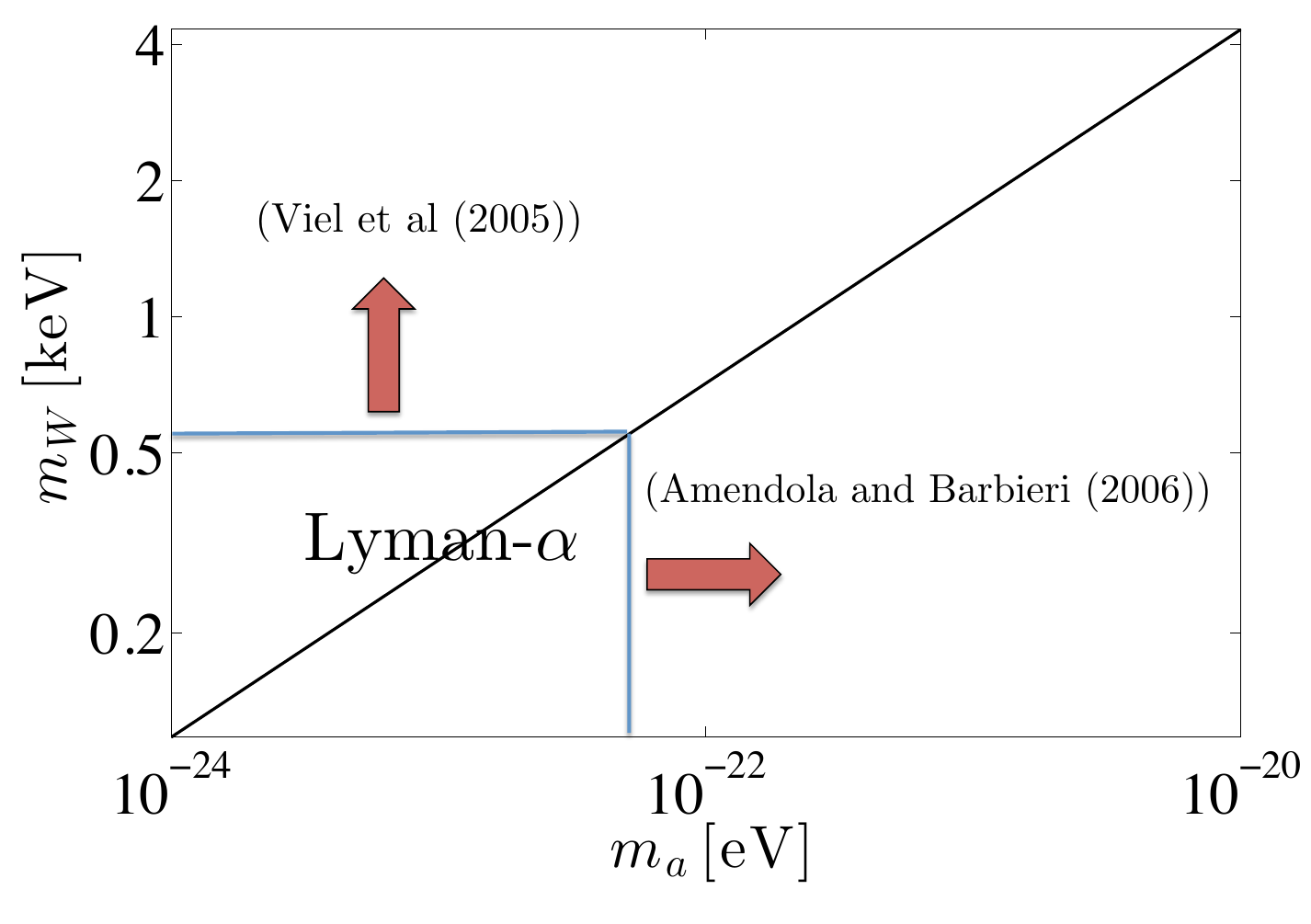}
\caption{Thermal relic warm dark matter mass in keV chosen to give the same transfer function half-mode, $k_m$ (Eq.~(\ref{eqn:km_def})), as a ULA, as a function of ULA mass in eV. We also show the Lyman-$\alpha$ forest constraints $m_W>0.55\text{ keV}$ of \citet{viel2005} corresponding to $m_a>5\times10^{-23}\text{ eV}$, consistent with \citet{amendola2005}.}\label{fig:mw_ma}
\end{figure}

Solving Eq.~(\ref{eqn:tk_wdm}) for the half-mode with WDM and using the fit with $\gamma=1.35$ to match $M_m(m_W)$ to $M_m(m_a)$, we plot $m_W(m_a)$ in Fig.~\ref{fig:mw_ma}. The power law relating them is $m_W\propto m_a^{0.39}$. Our matching to WDM mass applies to thermal relics like gravitinos, and on Fig.~\ref{fig:mw_ma} we show the constraint on thermal relics of $m_W>0.55\text{ keV}$ from Lyman-$\alpha$ forest data reported in \cite{viel2005}. This translates to a constraint on axion mass of $m_a>5\times 10^{-23}\text{eV}$, which is consistent with the Lyman-$\alpha$ constraints on ULAs reported in \cite{amendola2005}. The more recent Lyman-$\alpha$ constraints to WDM, such as \cite{viel2013} ($m_W\gtrsim 3.3 \text{keV}$) are much stronger, but there is no corresponding constraint to axions using this data to compare to.

Lyman-$\alpha$ constraints are sensitive to the exact shape of the transfer function: since mass goes as radius cubed, small differences between the transfer functions of ULA and WDM models will be amplified to larger differences in the associated mass scales. Lyman-$\alpha$ constraints also require careful calibration with hydrodynamical simulations \citep[as done in e.g.][]{viel2013}. Such simulations are available for CDM and WDM models, but not for ULAs, making the simple comparison of constraints by mass scale perhaps too naive.

\begin{figure}
$\begin{array}{@{\hspace{-0.3in}}l}
\includegraphics[scale=0.4]{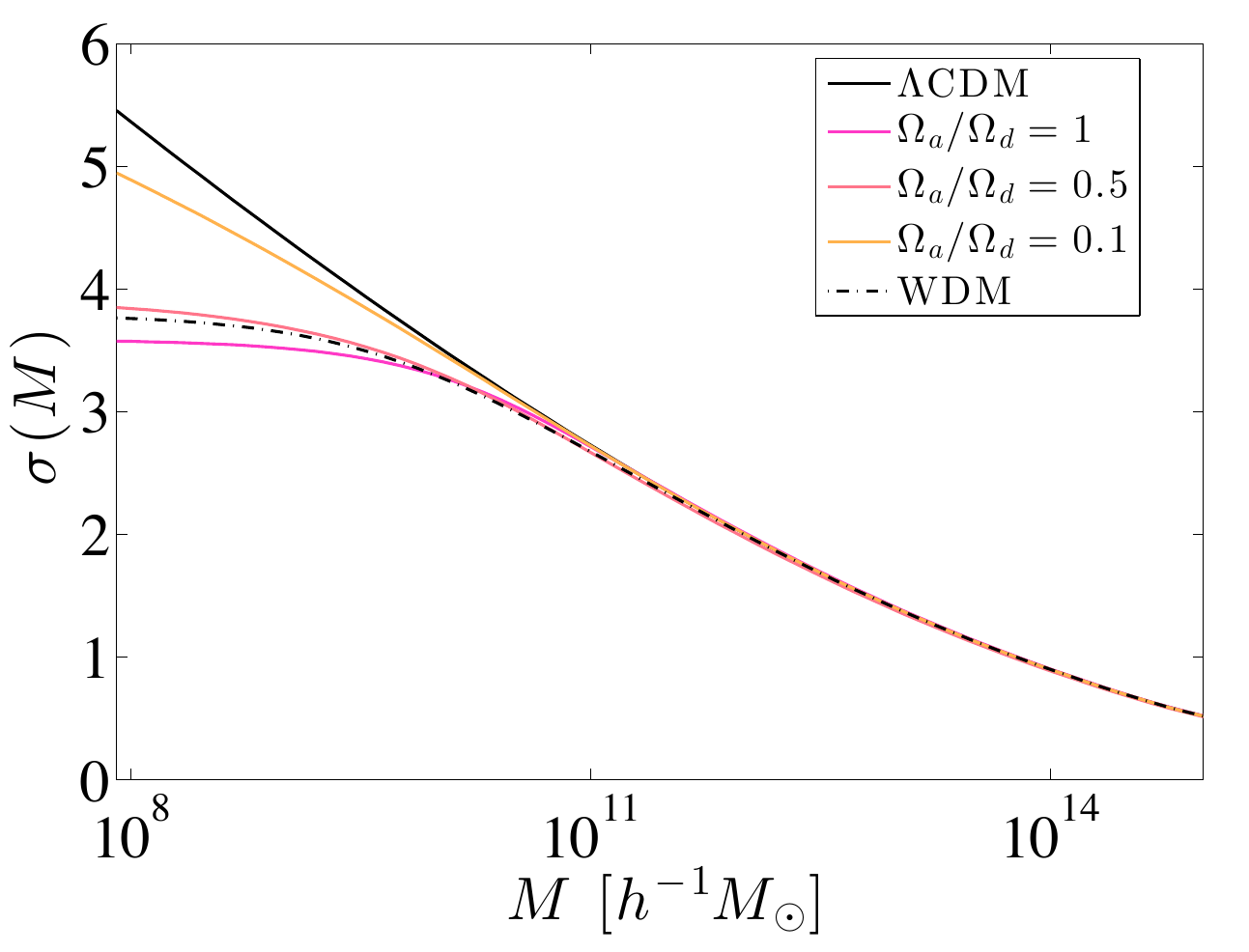} \\ [0.0cm]
 \end{array}$
\caption{Variance $\sigma(M)$ for $\Lambda$CDM, and aMDM with various $\Omega_a h^2$ at fixed total $\Omega_d h^2=0.112$ and axion mass $m_a=10^{-22}\text{ eV}$.}\label{fig:sigma_m_1e-22}
\end{figure}

The variance of the power spectrum, $\sigma(M)$, is computed by smoothing the power spectrum using a spherical top-hat window function of size $R$, and is done within \texttt{CAMB}:
\begin{align}
\sigma(R)^2&=\int_0^\infty \frac{dk}{k} P(k) W(k|R)^2 \, , \\
W(k|R)&=\frac{3}{(kR)^3}(\sin kR -kR \cos kR) \, .
\end{align}
In Fig.~\ref{fig:sigma_m_1e-22} we show the variance associated to the same models as in Fig.~\ref{fig:lin_tk}. The variance for the aMDM models varies little when the fraction is changed between $\Omega_a/\Omega_d=1$ and $\Omega_a/\Omega_d=0.5$, and is comparable to the associated WDM variance.

In the sections that follow we investigate the suppression of halo formation at and below $M_m$ in axion models in more detail.

\section{The Halo Mass Function}
\label{sec:hmf}

The MSP arises with CDM due to a larger expected number of low mass haloes than the number of low mass satellites observed in the Local Group (see e.g. \cite{primack2009} for a review).

To quantify this problem in various models we adopt the Press-Schechter (PS) approach \citep{press1974} to compute the abundance of halos of a given mass: the halo mass function (HMF). In the usual formalism this gives
\begin{align}
\frac{dn}{d\ln M} &= -\frac{1}{2} \frac{\rho_0}{M} f(\nu) \frac{d \ln \sigma^2}{d \ln M} \, , \\
\nu &\equiv \frac{\delta_c}{\sigma} \, ,
\end{align}
where $dn=n(M)dM$ is the abundance of halos within a mass interval $dM$. For the function $f(\nu)$ we use the model of \cite{sheth1999} (ST):
\begin{equation}
f(\nu)=A \sqrt{\frac{2}{\pi}} \sqrt{q} \nu (1+(\sqrt{q}\nu)^{-2p})\exp \left[  -\frac{q\nu^2}{2} \right] \, ,
\end{equation}
with parameters $\{A=0.3222,p=0.3,q=0.707\}$. The remaining ingredient in this approach is the critical overdensity, $\delta_c$, and what to do on mass scales $M<M_m$, both of which we now discuss.

\subsection{Mass Dependent Critical Density from Scale Dependent Growth}

In the case where all of the DM is made up of ULAs, as we saw in Fig.~\ref{fig:lin_tk}, there is no structure formed below $k_J$, and so we should expect no peaks in the density field, and thus no halos, below the mass scale $M_J$. However, applying the PS formalism described above with a constant barrier $\delta_c$ leads to a non-zero mass function for $M<M_J$. In the case of WDM this discrepancy is modelled in \cite{smith2011} by the addition of a smooth step in $dn/d\log M$ at $M=M_{fs}$. In the analytical results of \cite{benson2012} a much sharper cut-off was seen, and was attributed in part to a strong mass dependence in $\delta_c$, which was seen to increase rapidly below $M_{fs}$. A shallower cut-off was seen in the numerical results of \cite{angulo2013}. In the recent work of \cite{schneider2013} the cut-off due to free-streaming in WDM was investigated, and also found to be shallower than \cite{benson2012}. \cite{schneider2013} advocate a sharp $k$-space window function to match simulations and remove spurious structure thus providing the source of the cut-off: investigating different cut-offs and mass functions in aMDM will be the subject of a future work. 

In the absence of numerical simulations for ULA DM, or an existing treatment of the excursion set and spherical collapse in these models, one does not know what form the cut-off in the HMF near $M_J$ should take. In addition, for mixed dark matter models where the small-scale power is not entirely erased but only suppressed, one does not know how much (additional, \emph{ad hoc}) suppression to introduce. In this subsection we make a physically motivated argument for a mass dependent increase in $\delta_c$ at low $M$ that should account in some way for additional suppression in the HMF for $M<M_m$. 

Since we use results from \texttt{CAMB}, we take the overdensity $\delta$ to evolve with redshift, and in an Einstein-de Sitter (EdS) universe, take the critical overdensity to be fixed, $\delta_c=\delta_\text{EdS}\approx 1.686$. Alternatively, one can view the overdensity as being fixed, and take $\delta_c$ to evolve with redshift as $\delta_c(z)=D_0\delta_\text{EdS}/D(z)$ \citep{percival2000,percival2005}, which accounts for the growth between $z$ and $z=0$. The growth factor $D(z)$ is given by
\begin{equation}
D(z)=\frac{5 \Omega_m}{2 H(z)}\int_0^a \frac{da'}{(a' H(a')/H_0)^3} \, .
\end{equation}

In the aMDM model, there is \emph{scale-dependent growth} \citep[see e.g.][]{acquaviva2010,marsh2011b}, and we use this to model the change in $\delta_c$ with scale. For the relatively heavy axions we consider here the growth at the pivot scale $k_0=0.002 \,h \text{ Mpc}^{-1}$ is the same as in $\Lambda$CDM, while it is much smaller at $k>k_m$. We take $\delta_c(k)$ at $z=0$ to be altered by an amount $D(k_0)/D(k)$, and normalise by the same ratio in $\Lambda$CDM (to take account of the small amount of scale-dependent growth there). In the interests of simplicity, we will only be concerned with examples of the HMF at $z=0$ and take $\delta_c(z=0,k=k_0)=\delta_\text{EdS}$, which is good to within a few percent for $\Lambda$CDM \citep{percival2005}\footnote{A more advanced treatment of spherical collapse in coupled quintessence cosmologies in \cite{tarrant2011} found that even in $\Lambda$CDM $\delta_c(z=0)$ can differ from $\delta_\text{EdS}$ by more than this amount.}. At redshift $z=0$ our model takes
\begin{align}
\delta_c(k)&=\mathcal{G}(k)\delta_\text{EdS} \, , \\
\mathcal{G}(k) &\coloneqq \frac{D(k)_{\Lambda\text{CDM}}}{D(k)_\text{aMDM}} \, . \label{eqn:g_def}
\end{align}

Two, not entirely unrelated, issues arise with this model when trying to extract the growth from a Boltzmann code. The first is that to use this model we must disentangle growth from transfer function, which is by definition somewhat problematic in the case of scale-dependent growth. Defining the transfer function as the piece which depends solely on $k$ this can only be done with the logarithmic derivative $d\log \delta/d\log a=d\log D/\ \log a$, which does not give us the absolute value at $z=0$ that we seek. We take a more practical definition suited to numerical computation. In $\Lambda$CDM, the transfer function freezes in somewhere around the decoupling epoch \citep{eisenstein1998}, when matter domination is total. This provides a definition of the scale-dependent growth at $z=0$ which is easily accessible from a numerical solution for $\delta(k,z)$, normalised such that $D(k=k_0)=1$:
\begin{equation}
\frac{D(k)}{D_0}\coloneqq \frac{\delta(k,0)}{\delta(k,z_h)}\frac{\delta(k_0,z_h)}{\delta(k_0,0)} \, ,
\end{equation}
where $z_h$ is chosen so that in $\Lambda$CDM the transfer function has frozen in, and $k_0\ll k_m$ is the pivot scale. Using \texttt{CAMB}, we find $z_h\approx 300$ works well. We then use exactly the same definition to set the scale of $D(k)$ in the aMDM case.

\begin{figure}
\begin{center}
$\begin{array}{@{\hspace{-0.3in}}l}
\includegraphics[scale=0.35]{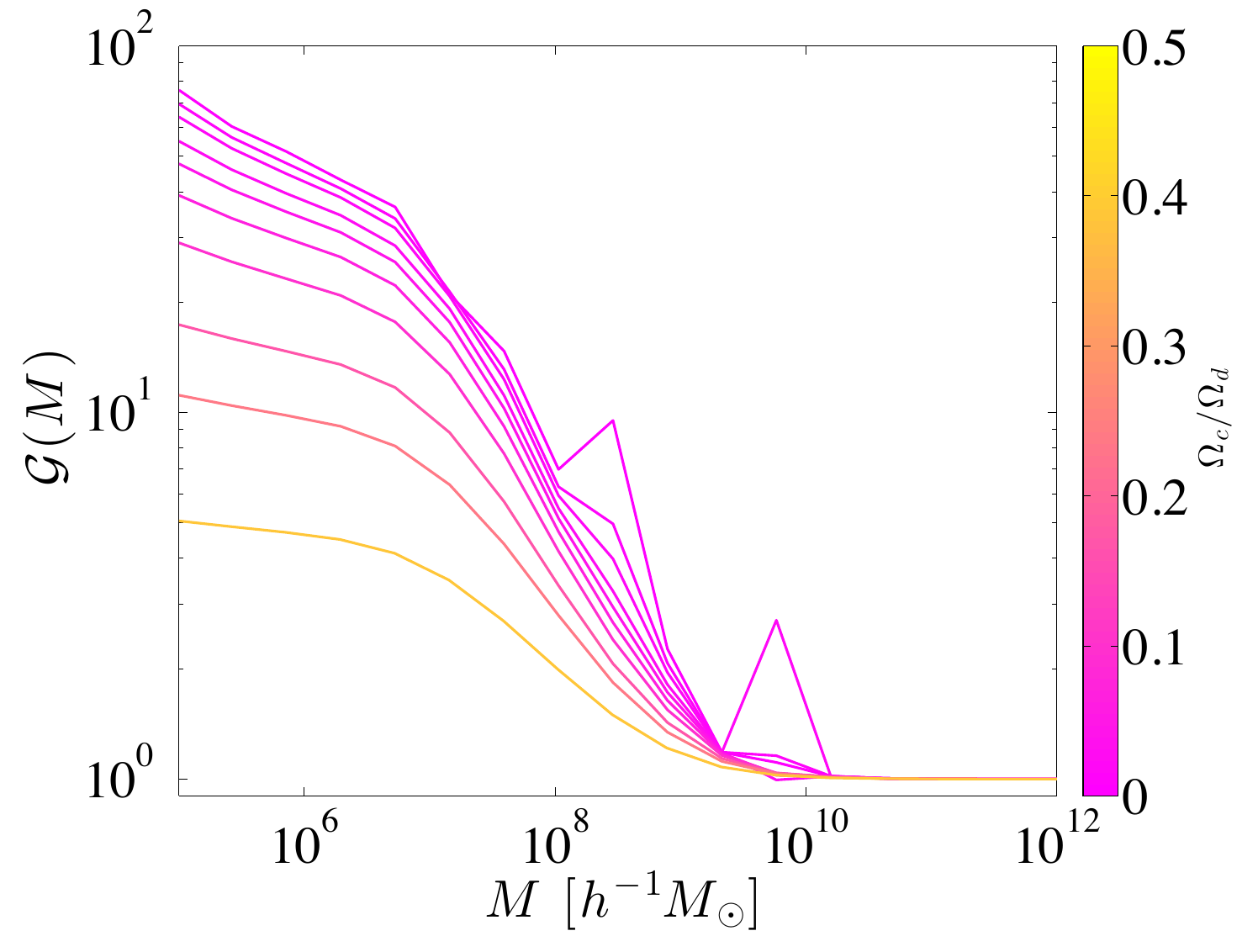} \\ [0.0cm]
 \end{array}$
 \caption{The mass dependent critical density from scale-dependent growth, Eq.~(\ref{eqn:g_def}), is given by $\delta_c(M)=\mathcal{G}(M)\delta_\text{EdS}$. We show $\mathcal{G}(M)$ for aMDM with various $\Omega_c/\Omega_d$ at fixed total $\Omega_d h^2=0.112$ and axion mass $m_a=10^{-22}\text{ eV}$. The spikes at low fraction are due to BAO distortions and numerical instability defining scale-dependent growth via a ratio.}\label{fig:gm_1e-22}
\end{center}
 \end{figure}

Scale dependent growth causes the mass dependent critical density to increase below $M\approx M_m$. Fig.~\ref{fig:gm_1e-22} shows $\mathcal{G}(M)$ for these models. There is the obvious trend that $\mathcal{G}(M)$ decreases with increasing CDM fraction. 

The second issue is that if the axions completely dominate the matter density then the overdensity will become vanishingly small for $k\gg k_m$ even at high redshift, and so we are faced with the problem of dividing zero by zero to set the scale of $D(k)$. This is a numerical precision problem and, when combined with BAO distortions, leads to the spikey/oscillatory behaviour of $\mathcal{G}(M)$ for $\Omega_c/\Omega_d\lesssim 0.01$ in Fig.~\ref{fig:gm_1e-22}. 
%This numerical problem is only an issue when axions make up a very large fraction of the DM, $f_{ax}\gtrsim 0.95$: elsewhere the suppression of $P(k>k_m)$ is not so severe as to approach machine precision in the ratio defining $D(k)$. In any case, when axions make up all of the DM, as mentioned above, we might expect usual arguments to break down, and more careful treatments to be necessary.

\subsection{HMF Results}

\begin{figure}
\begin{center}
$\begin{array}{@{\hspace{-0.2in}}l}
\includegraphics[scale=0.35]{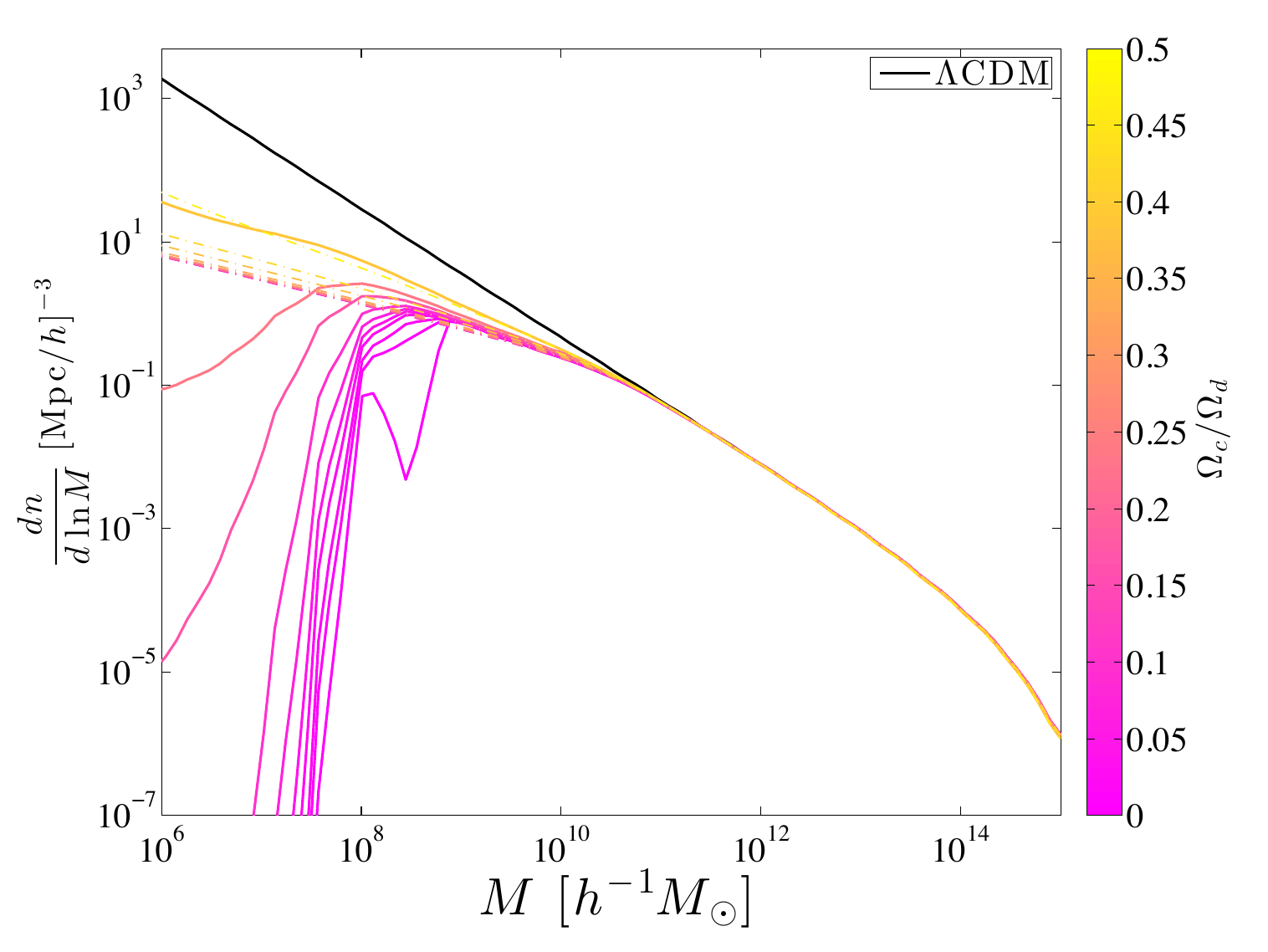} \\ [0.0cm]
 \end{array}$
\caption{Halo mass function computed directly from \texttt{\texttt{CAMB}} for $\Lambda$CDM, and aMDM with various $f_c=\Omega_c/\Omega_d$ at fixed total $\Omega_d h^2=0.112$ and axion mass $m_a=10^{-22}\text{ eV}$. Solid lines have $\delta_c(M)$ while dashed lines have $\delta_c=\delta_\text{EdS}$. At $f_c=0.5$ the cut-off is no longer present in the mass range shown, the difference between $\delta_\text{EdS}$ and $\delta_c(M)$ having largely vanished. The spikes at low fraction are due to BAO distortions and numerical instability defining scale-dependent growth via a ratio.}\label{fig:hmf_deltac_m}\end{center}
 \end{figure}

In Fig.~\ref{fig:hmf_deltac_m} we plot the HMF with a fixed axion mass of $m_a=10^{-22}$eV for a variety of values of the axion density, with fixed total $\Omega_d h^2=0.112$.  We see suppression of the mass function beginning at $M_m$, with the amount of suppression increasing and the asymptotic slope of the mass function decreasing as we raise the axion density. We show results taking $\delta_c$ fixed, and those with mass dependent $\delta_c(M)$, modelled for as above. 
%The models with small $f_c$ shown in Fig.~\ref{fig:hmf_deltac_m} suppress the linear power spectrum to very similar degrees. This is seen in the HMF from the dashed lines in Fig.~\ref{fig:hmf_deltac_m} being very close together. Suppression of the variance begins around $M_m\approx 10^{10}\, h^{-1}M_\odot$ and quickly flattens out towards lower masses.

The introduction of scale-dependent growth via $\mathcal{G}(M)$ in Fig.~\ref{fig:hmf_deltac_m} causes the HMF to be sharply cut off at around $M\approx 10^8 \, h^{-1}M_\odot\approx 0.01 M_m$ with large axion fraction. This is in agreement with the cut-off of \cite{smith2011} and the numerical results of \cite{angulo2013} for WDM: the HMF falls below its $\Lambda$CDM value at the half-mode mass, but only cuts off completely at a lower mass, intermediate between the Jeans (free-streaming for WDM) mass and the half-mode mass. By considering fragmentation of proto-halo objects formed in WDM cosmologies \cite{angulo2013} found a smoother cut-off in the HMF than the sharp cut-off of \cite{benson2012} coming from analytic results. By the time we reach the Jeans scale of $M_J\approx 1.1 \times 10^7 \, h^{-1}M_\odot$ the mass function for $\Omega_a/\Omega_d=1$ is vanishingly small, more than eight orders of magnitude below its $\Lambda$CDM value.
\begin{figure}
\begin{center}
$\begin{array}{@{\hspace{-0.4in}}l}
\includegraphics[scale=0.45]{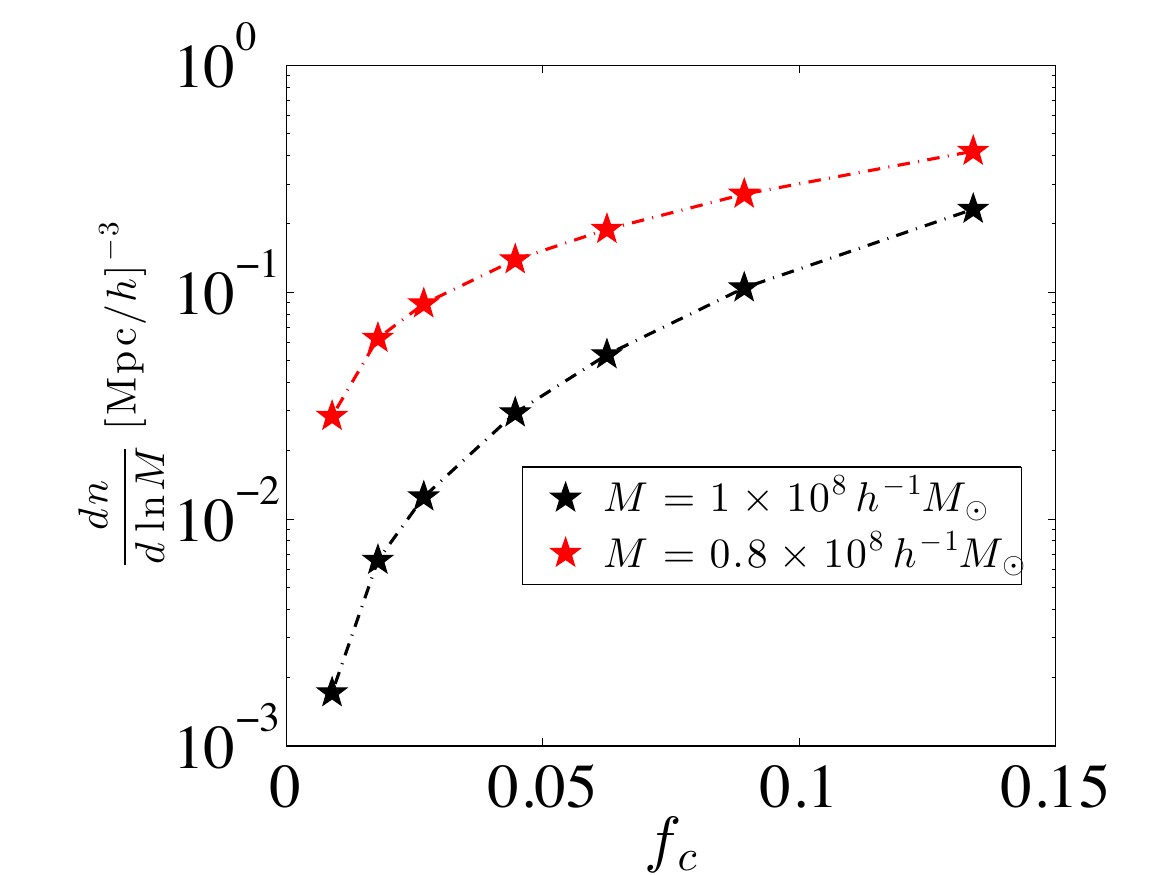} \\ [0.0cm]
 \end{array}$
 \caption{The HMF evaluated at $M=1\times 10^8 \, h^{-1}M_\odot\approx 0.01 M_m$ and $M=0.8\times 10^8 \, h^{-1}M_\odot$ for aMDM as a function of $f_c=\Omega_c/\Omega_d$ at fixed total $\Omega_d h^2=0.112$ and axion mass $m_a=10^{-22}\text{ eV}$. The HMF decreases rapidly below the cut-off, at around 1\% of the half-mode mass. Varying the CDM fraction between $1$ and $15\%$ can change the value of the HMF at the cut-off by two orders of magnitude.}\label{fig:hmf_cutoff_1e-22}
\end{center}
 \end{figure}

In Fig.~\ref{fig:hmf_cutoff_1e-22} we plot the HMF evaluated at various masses near $M=0.01 M_m$ as a function of $f_c=\Omega_c/\Omega_a$ at low $f_c$ to investigate the effect of a small admixture of CDM on the value of the mass function at the cut-off. Varying $f_c$ between $1$ and $15\%$ can change the value of the HMF near the cut-off by two orders of magnitude. The small admixture of CDM can help an aMDM model form dwarf halos near the HMF cut-off.

The low values of $f_c\lesssim 0.13$, as we will see in Section~\ref{sec:two_component_profile}, are those relevant for core formation with aMDM. At larger values of $f_c$ approaching the equally mixed DM $f_c=0.5$ the sharp cut-off in the HMF has vanished, although it is still significantly reduced compared to $\Lambda$CDM. The large admixture of CDM, if the need for cores is foregone, still remains relevant to the MSP and introduces no potentially problematic cut-off in the HMF.

Scale dependent growth in aMDM induces a cut-off in the HMF similar to the cut-off observed in numerical simulations of WDM. In \cite{angulo2013} the cut-off in WDM simulations could be fit by introducing non-spherical filtering to compute $\sigma (R)$. By assigning masses to radii differently for WDM compared to CDM after accounting for formation of halos by fragmentation this cut-off was made less severe. In order to discuss the assignment of masses to halos in aMDM we now move on to model the halo density profile and its normalisation.

%There are two key conclusions to be drawn from the preceding discussion. Firstly, the mass dependence of the critical density, and thus the cut off in the HMF only, appears strongly at masses slightly \emph{below} the linear Jeans scale. Secondly, at this scale the value of the HMF is strongly dependent on the exact composition of the DM. Both of these conclusions will have a part to play in the resolution of the \emph{Catch 22} using aMDM.

%The cut off in the HMF and thus the \emph{Catch 22} for WDM, is inferred from the mass dependent critical density fit from N-body simulations, and these simulations are specific to the case of WDM. In Ref.~\citep{benson2012} $\delta_c(M)$ for WDM begins to rise quite close to $M_J$, and rises steeply as the mass decreases, increasing by over two orders of magnitude at $M/M_J=0.01$. This is an earlier and sharper rise than we observe in the model for $\mathcal{G}(M)$ based on scale-dependent growth in aMDM, and will have consequences for the \emph{Catch 22}. Without an N-body simulation for aMDM, we will take the model based on scale-dependent growth to be correct.
\section{Halo Density Profile}
\label{sec:density_profile}

The \emph{Catch 22} \citep{maccio2012} of solving the CCP with WDM is that the WDM particle mass required to introduce a core of sufficient size in a dwarf galaxy serves to cut off the HMF at exactly the mass of the dwarf, so that it is never formed. At the same time, WDM allowed by constraints from LSS does not form cores of relevant (kiloparsec) size. In order to ascertain whether the \emph{Catch 22} applies to aMDM, or indeed to the case of pure axion DM, we must must model the expected core size.

We follow \cite{hu2000} and associate a core size to the Jeans scale \emph{within the halo}, $r_{J,h}$, below which the density will be assumed constant\footnote{See also \cite{arbey2001} and \cite{arbey2003} who studied the effect of scalar DM of mass $m_a\approx 10^{-23}$ eV on galaxy rotation curves in the presence of baryons, and core formation in the Bose condensate. Yet another model of cores is considered by \cite{bernal2003}. None of these models consider the altered cosmology and structure formation.}. The Jeans scale within the halo is related to the linear Jeans scale, $r_J$, by scaling the energy density in Eq.~(\ref{eqn:hu_jeans_def})
\begin{equation}
r_{J,h}=\left(\frac{\rho_0}{\rho(r_{J,h})}\right)^{1/4}r_J \, .
\label{eqn:jeans_halo}
\end{equation}
Thus we can determine the linear Jeans scale (and so the ULA mass) necessary to provide a given core size inside a dwarf halo, if we know the external profile $\rho (r)$. The assumption inherent in Eq.~(\ref{eqn:jeans_halo}) is that the coherent effects in the scalar field giving rise to the Jeans scale survive in the non-linear regime when mode mixing becomes important and the linear derivation of the sound speed in Eq.~(\ref{heuristic_cs}) may break down. $N$-body/lattice simulations of the axion field are needed to test this assumption.

Assuming that collapse occurs as in $\Lambda$CDM, \cite{hu2000} computed $\rho_0/\rho (r_{J,h})$ and found that a core of size $r_{J,h}\sim 3.4 \text{ kpc}$ is obtained in a dwarf halo of mass $10^{10}M_\odot$ for an axion of mass $m_a = 10^{-22}\text{ eV}$. As we have seen, the HMF for such a ULA is only cut off for $M\lesssim 10^8 h^{-1}M_\odot$ suggesting that axions do not suffer the \emph{Catch 22} of WDM.

In the following section we address this is in a more detailed model of aMDM. Firstly, we compute halo parameters with the pure ULA variance, normalise our cored halo profile, and find the relationship between ULA mass and core size in a representative Milky Way satellite. The picture that emerges is qualitatively the same as \cite{hu2000}, but quantitatively different. Secondly, we extend this picture to a two-component profile and ask whether cores can be maintained as a small admixture of CDM is added.

\subsection{The NFW Profile}

For the external profile, $\rho (r)$, outside of the Jeans scale where the ULA behaves as CDM, we use the universal radial density profile of \cite{navarro1997} (hereafter, NFW):
\begin{equation}
\frac{\rho(r)}{\rho_0}=\frac{\delta_{\text{char}}}{(r/r_s)(1+r/r_s)^2} \, ,
\label{eqn:nfw_profile}
\end{equation}
where the scale radius $r_s=r_{200}/c$, with $r_{200}$ the virial radius, $c$ the concentration parameter, and $\delta_{\rm char}$ the characteristic density. 

The characteristic density is assumed to be proportional to the density of matter in the universe at the collapse redshift of the halo, $z_{\text{coll}}$. The definition of $z_{\rm coll}(M)$ is fixed for NFW and follows from Press-Schechter \citep[see also][]{lacey1993} :
\begin{equation}
\text{erfc}\left(\frac{\delta_{c,\text{EdS}} (D(z_{\text{coll}})^{-1}-1)}{\sqrt{2(\sigma^2(fM)-\sigma^2(M))}} \right)=\frac{1}{2}\, ,
\label{eqn:zcoll_nfw}
\end{equation}
The NFW profile  is fit with $f=0.01$. As above we work in the convention where $\delta_c$ is constant but the overdensities themselves evolve with linear growth factor $D(z)$. The characteristic density is then 
\begin{equation}
\delta_{\rm char}=C \Omega_m (1+z_{\rm coll})^3 \, ,
\end{equation}
where $C=3.4 \times 10^3$ is fit by NFW to simulations of CDM, which should match axion DM above the Jeans scale.

The virial radius is defined as the radius at which the average enclosed density is $200$ times the mean density, in terms of the halo mass $M$ at redshift $z=0$ it is given by:
\begin{align}
r_{200}(M,z)=&\left( 200 \frac{4}{3}\pi \right)^{-1/3} \left(\frac{\rho_0}{h^2 \text{kpc}^{-3} M_\odot}\right)^{-1/3} \nonumber \\
	& \left( \frac{M_{200}}{h^{-1}M_\odot} \right)^{1/3}\, h^{-1}\text{ kpc} \, .
\end{align}
For the NFW profile the concentration and scale radius, with the correct choice of $C$, are defined such that $M=M_{200}$. For the cored profile that we discuss below the scale radius of the external NFW profile does not have the same relationship with the true virial radius, and we normalise separately for $M_{200}$. 

Finally, the concentration is defined from the characteristic density, $\delta_{\text{char}}$, by
\begin{equation}
\delta_{\text{char}}=\frac{200}{3}\frac{c^3}{\ln (1+c)-c/(1+c)} \, .
\end{equation}

The definition of $z_{\rm coll}$ in Eq.~(\ref{eqn:zcoll_nfw}), and hence the concentration defined from it will go to zero for a variance that flattens out at low masses, as is the case for aMDM with small CDM fraction. The lower concentration of low mass halos in comparison to $\Lambda$CDM will be relevant for MFP, which we discuss in Section~\ref{sec:tbtf}. Since $z_{\rm coll}$ is also lower, in Section~\ref{sec:collapse} we discuss the collapsed mass fraction and potential conflicts with observations of high-redshift galaxies.
%When we consider whether the flattening of the variance in axion models (Fig.~\ref{fig:sigma_m_1e-22}) may introduce conflicts with the number of observed high redshift galaxies we will want to consider $3\sigma$ rare fluctuations, and as such in Section~\ref{sec:collapse} will define the collapse redshift with $F_{\rm coll}=0.005$ instead. 
%The definition Eq.~(\ref{eqn:zcoll_nfw}) follows from the halo mass function and it immediately gives another manifestation of the \emph{Catch 22}. When the variance, $\sigma^2(M)$, is truly flat below $M_m$, which is approximately the case with all the DM in axions, then the denominator in Eq.~(\ref{eqn:zcoll_nfw}) goes to zero, and the halo never collapses. From Fig.~\ref{fig:sigma_m_1e-22} we see that this is only mildly alleviated with the introduction of a small amount of CDM and is only exacerbated by including scale-dependent growth which increases the numerator. In aMDM, if the dwarves collapse in this way, they will only be formed at low redshift.

\subsection{Halo Jeans Scale For Pure Axion DM}
\label{sec:jeans_and_norm}

In this subsection we consider the core size, and normalisation of halos in a pure ULA dark matter model. We assume collapse occurs as in $\Lambda$CDM, with $D(z)$, but use the axion variance, $\sigma(M)$. 

For definiteness, we consider halos with the simplest possible cored profile
\begin{align}
\rho_{\rm core}(M,r)=&\theta(r-r_{J,h}(M)) \rho_{\rm NFW}(M,r)\nonumber \\
&+\theta(r_{J,h}(M)-r)\rho_{\rm NFW}(M,r_{J,h}(M)) \, ,
\label{eqn:simple_core}
\end{align} 
where $\theta (x)$ is the Heaviside function, although much of what we say below will apply to any cored profile with core radius $r_c=r_{J,h}$ fixed by Eq.~(\ref{eqn:jeans_halo})\footnote{Other cored profiles are studied in e.g. \cite{zavala2012} for self-interacting DM.}. In particular, the choice of a Heaviside function introduces sharp transitions into the density profile, and as such is only for illustration. The external NFW profile is consistent with what is observed in the WDM simulations of \cite{maccio2012}. 

In order to find the Jeans scale within a halo we must solve Eq.~(\ref{eqn:jeans_halo}) for an NFW profile with external profile normalisation fixed at $M_s$ (the `scale mass'), and shape fixed by scale radius $r_s(M_s)=r_{200}(M_s)/c(M_s)$ to find $r_{J,h}(M_s)$. This is \emph{not} the Jeans scale within a halo of mass $M=M_s$: the mass $M_s$ is the mass that an equivalent NFW profile would have. We will discuss normalisation of $M$ shortly.

We use the variance for the axion model to compute our NFW halo parameters: using the $\Lambda$CDM variance, the halo Jeans scales with fixed $M_s$ will be different. With low $M_s$ relative to $M_m$ the concentration is lower when the correct variance is used, which causes the Jeans scale to be smaller by the increase in scale radius relative to $r_{200}$. On the other hand the Jeans scale within intermediate and high mass objects is found to be larger with the correct variance. For example, with $m_a=10^{-21}\text{ eV}$ the shift in halo Jeans scale inside dwarf galaxies can be of order $0.1\, h^{-1}\text{kpc}$.

\begin{figure}
\begin{center}
$\begin{array}{@{\hspace{-0.3in}}l}
\includegraphics[scale=0.32]{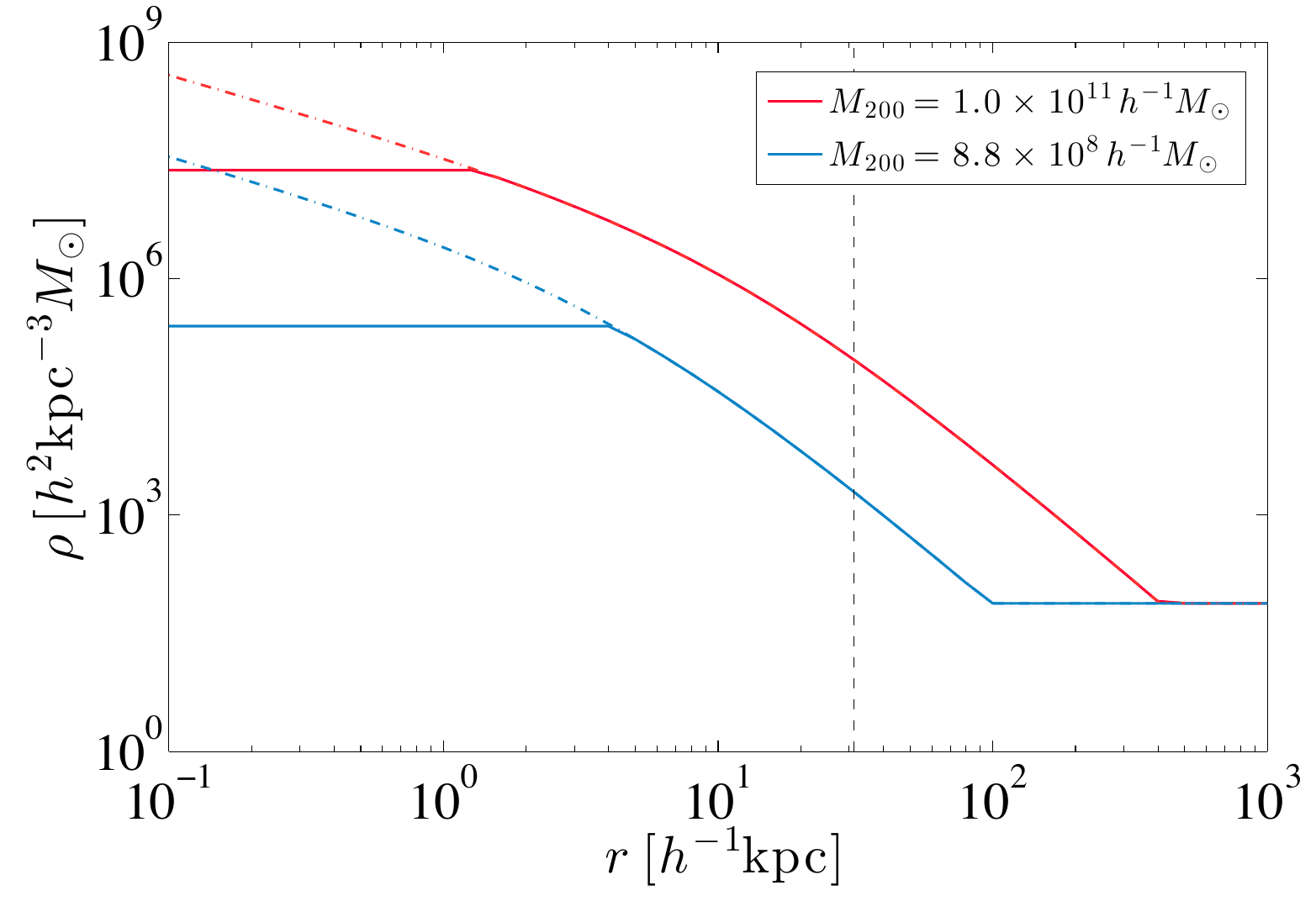} \\ [0.0cm]
 \end{array}$
 \caption{Halo profiles of Eq.~(\ref{eqn:simple_core}) (solid lines) with various $M_{200}$ compared to their parent NFW profiles of mass $M_s\approx M_{200}$ (dot-dashed lines). The axion mass is $m_a=10^{-22}\text{ eV}$ and we show the linear Jeans scale, $r_J=31.2 \, h^{-1}\text{ kpc}$ (vertical dashed line). As long as halo becomes overdense outside of $r_J$ it can continue to be overdense inside until it reaches the halo Jeans scale, $r_{J,h}$ satisfying Eq.~(\ref{eqn:jeans_halo}). More massive halos are more dense at $r_J$ and the halo Jeans scale is smaller. No profiles form with $M_s<M_{\rm low}$ when the NFW parent has not become overdense outside of $r_J$ (although they may form by fragmentation).}\label{fig:one_component_core_profile}
\end{center}
 \end{figure}

The Jeans scale decreases in higher density environments and therefore the positive real solution of Eq.~(\ref{eqn:jeans_halo}) is a monotonically decreasing function of $M_s$. At low enough $M_s$, then, one finds $r_{J,h}>r_J$. This cannot be a physical solution. Solutions to Eq.~(\ref{eqn:jeans_halo}) giving $r_{J,h}>r_J$ can only occur when $\rho/\rho_0<1$, which represents a void and not a halo\footnote{Solutions to Eq.~(\ref{eqn:jeans_halo}) with $r_{J,h}>r_J$ also reminds us of another overlooked feature of pure ULA dark matter: not only are halos cored, having a maximum density, but voids will also be `cored', having a maximum underdensity. This suggests other possible probes/virtues of ULAs \citep{peebles2010}, but we will not consider this possibility further here.}. This break at $r_J=r_{J,h}$ will occur at a certain mass, $M_{\rm low}(M,r_J)$, which is a function of the linear Jeans scale. It is found by solving
\begin{equation}
\rho_{\rm NFW}(M_{\rm low},r_J)=\rho_0 \, .
\end{equation}
This tells us that no cored halos with normalisation $M_s<M_{\rm low}$ exist for fixed $r_J$. Cored halo profiles of various masses are shown in Fig.~\ref{fig:one_component_core_profile} and compared to their parent NFW profiles, with the linear Jeans scale shown for scale. It is clear that no halos should form that do not already become overdense outside $r_J$. In the example, with $r_J=31.2 \, h^{-1}\text{ kpc}$, this implies that halos with $M_s=4\times 10^8 h^{-1}M_\odot$ only just form. 

 % For $M_s<M_{\rm non-vir}$ the maximum density of the halo never exceeds $200 \rho_0$ (horizontal dashed line): we refer to these halos as non-virialised as they cannot have a mass $M_{200}$ associated to them.
%\begin{figure}
%\includegraphics[scale=0.8]{jeans_halo_mass.pdf}
%\caption{Contour plot of the halo Jeans scale, $r_{J,h}(\log_{10}M_s, \log_{10}(r_J/\,h^{-1}\text{ kpc}))$ within a halo of NFW mass $M_s$, in units of the linear Jeans scale, $r_J$. Notice that, despite the Jeans scale increasing in high density environments, the solutions to Eq.~(\ref{eqn:jeans_halo}) are monotonically decreasing with $M_s$. When $r_{J,h}>r_J$ we have $\rho<\rho_0$, i.e. this is not valid in an overdense region like a halo. This occurs at $M_s=M_{\rm low}$ and signals the fact that no halos with external scale $M_s<M_{\rm low}$ are formed. Shaded regions of the plane have $r_{J,h}(r_J,M_s)/r_J\leq 1$ and represent regions of halo formation. Note that this region is below the pink dashed line $M_J$.}\label{fig:jeans_halo_mass}
%\end{figure}

Given that no halos form with external scale $M_s<M_{\rm low}$, what is the minimum halo mass in this model? A halo is normalised to mass $M_{200}$ by the integral out to the virial radius, $r_{200}$
\begin{equation}
4 \pi \rho_0 \int_0^{r_{200}} r^2 \rho_{\rm core} (M,r) \,\mathrm{d}r=M_{200}\equiv 200 \rho_0 \frac{4}{3}\pi r_{200}^3 \, .
\label{eqn:simple_core_norm}
\end{equation}
This defines $M_{200}(M_s)$. Since the Jeans scale within halos is a monotonically decreasing function of $M_s$, at some scale mass $M_{\rm non-vir}$ the solution for $r_{200}$ will occur when $\rho_{\rm NFW}(M_{\rm non-vir},r_{J,h}(M_{\rm non-vir}))=200 \rho_0$. Halos with scale mass $M_s<M_{\rm non-vir}$ will not be virialised in the sense that their average density never exceeds $200$ times the background density. In order to assign mass to these halos one cannot use $M_{200}$. The total enclosed mass is found by integrating out to $\rho(r_1)=\rho_0$. With this alternative definition of mass it is clear that the lowest mass object formed at $M_s=M_{\rm low}$ is at exactly the Jeans mass, $M_J$.

Halos with $M<M_{200}(M_{\rm non-vir})$ will also need to have masses assigned to radii coming form the linear filtering in $\sigma (M|R)$ in the HMF in a different manner than is applicable to CDM. This is accounted for by fits to simulations in \cite{angulo2013} and is partly responsible for the less severe cut-off in the HMF found in that work. By comparison, if we made the same assignment in the HMF with pure axion DM, the cut-off found by introducing scale-dependent growth might also become less severe.

We typically find that, for $M_s>M_{\rm non-vir}$, $r_{200}$ for the cored profile is approximately the same as for the parent NFW, and that $M_{200}\approx M_s$. Approaching $M_{\rm non-vir}$, $M_{200}$ drops below $M_s$.

\begin{figure}
$\begin{array}{@{\hspace{-0.15in}}l}
\includegraphics[scale=0.35]{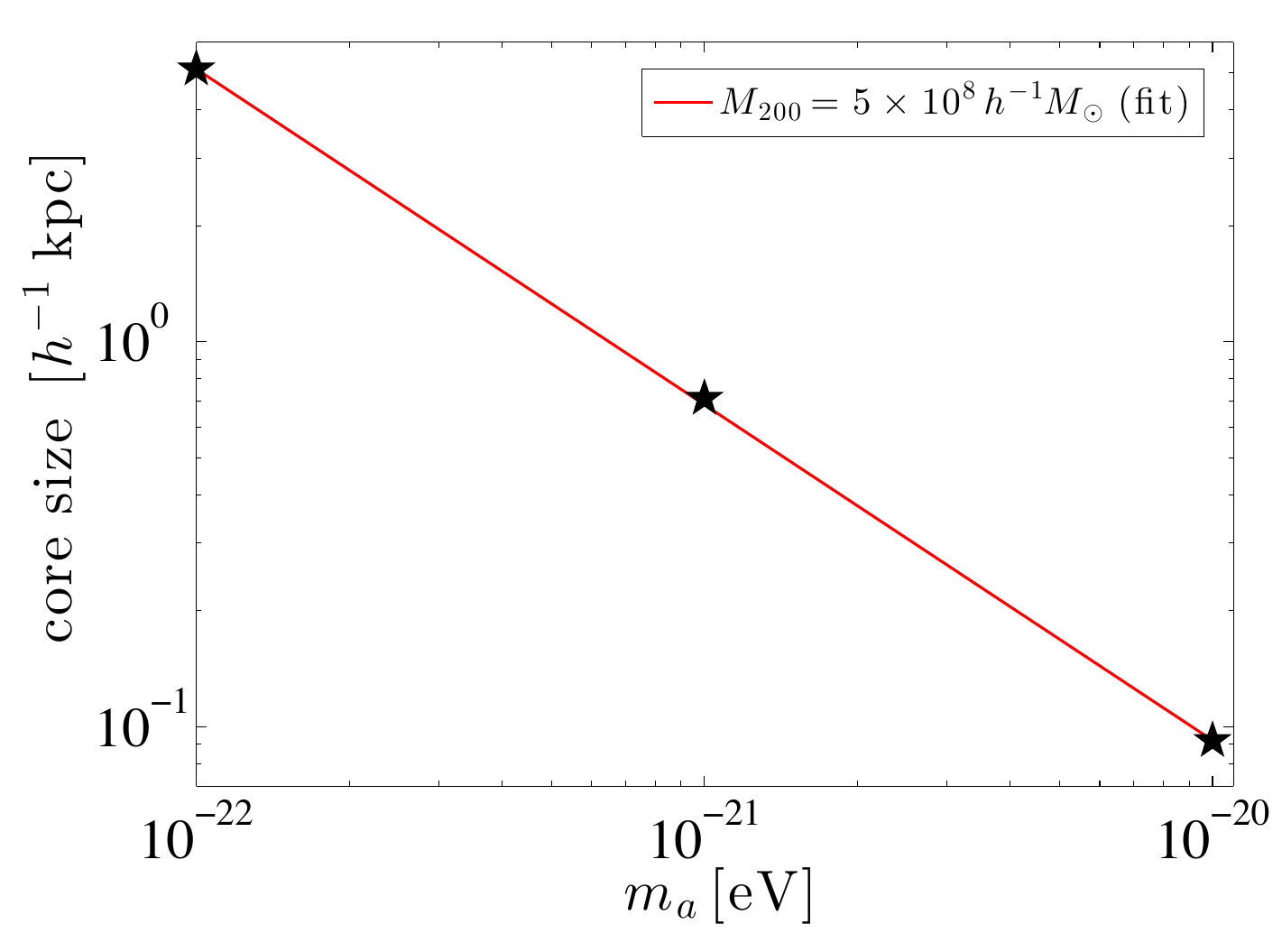} \\ [0.0cm]
 \end{array}$
\caption{The expected core size in a typical Milky Way satellite of mass $M_{200}=5\times 10^8 \, h^{-1}M_\odot$ as a function of axion mass, $m_a$. The relationship is well fit by a simple power law, shown as $m_a^{-0.87}$, but due to the dependence of the concentration on $m_a$ this is not the scaling of $r_J$ with $m_a$. The core size is fairly significant, $\gtrsim 0.1 \, h^{-1}\text{ kpc}$, across the entire mass range, which demonstrates that axions satisfying all large scale structure constraints can provide a potentially viable resolution of the cusp-core problem in CDM halos.}\label{fig:rjh_axionmass}
\end{figure}

Finally, having normalised our halos, we show in Fig.~\ref{fig:rjh_axionmass} the core size expected in the typical Milky Way dwarf galaxy of mass $M_{200}=5 \times 10^8\, h^{-1}M_\odot$ as a function of axion mass for the heavier axions $m_a \geq 10^{-22}\text{ eV}$ allowed by the relevant Lyman-$\alpha$ constraints. The core size is well fit by a single power law in axion mass. The best fit power law has $r_{J,h}\sim m_a^{-0.87}$, while the power law given in \cite{hu2000} is $r_{J,h}\sim m_a^{-2/3}$. \cite{hu2000} used approximate formulae to solve Eq.~(\ref{eqn:jeans_halo}) and do not give the dependence of the concentration on axion mass. The dependence of the concentration on axions mass comes from computing NFW parameters with the correct variance and leads to the different power law.

We find that there is a considerable core size for all masses considered. In particular, even our heaviest axion with $m_a=10^{-20}\text{ eV}$ has a core size of $r_{J,h}=0.1\, h^{-1}\text{ kpc}$. This heaviest axion has a characteristic mass scale of $M_m=10^8\, h^{-1}M_\odot$ and would not affect the formation rate of these dwarf galaxies in any dramatic way. This demonstrates that pure ULA dark matter does not suffer from the \emph{Catch 22} of WDM: ULAs allowed by even the most stringent large scale structure constraints, which would barely affect the HMF at Milky Way satellite masses, can still give significant cores to dwarf galaxies.

\subsection{A Mixed Dark Matter Halo Profile}
\label{sec:two_component_profile}

Having understood the simple cored profile in pure axion DM, we now move on to consider the two-component dark matter subclass of aMDM (see \cite{medvedev2013} and references therein for recent work on two-component DM halos). In this subsection we continue to assume $\Lambda$CDM growth. We study axion mass $m_a=10^{-22}\text{ eV}$ and a benchmark halo with $M_{200}=5 \times 10^{8} \, h^{-1}M_\odot$.

Again, we take the halo density profile to be NFW outside of the halo Jeans scale, while inside the halo Jeans scale we take the axion component to be smooth. Inside this smooth background we then assume the CDM component of the DM collapses as usual and forms its own NFW halo. Because the axion component is smooth inside this radius, we superpose the profiles and therefore below the Jeans scale the ratio of axions to CDM is not constant but decreases at small radius. This is in agreement with the simulations of mixed cold plus warm DM in \cite{anderhalden2012}. The assumed 2-component profile is given in Appendix~\ref{appendix:normalisation}.  

\begin{figure}
\begin{center}
$\begin{array}{@{\hspace{-0.15in}}l}
\includegraphics[scale=0.35]{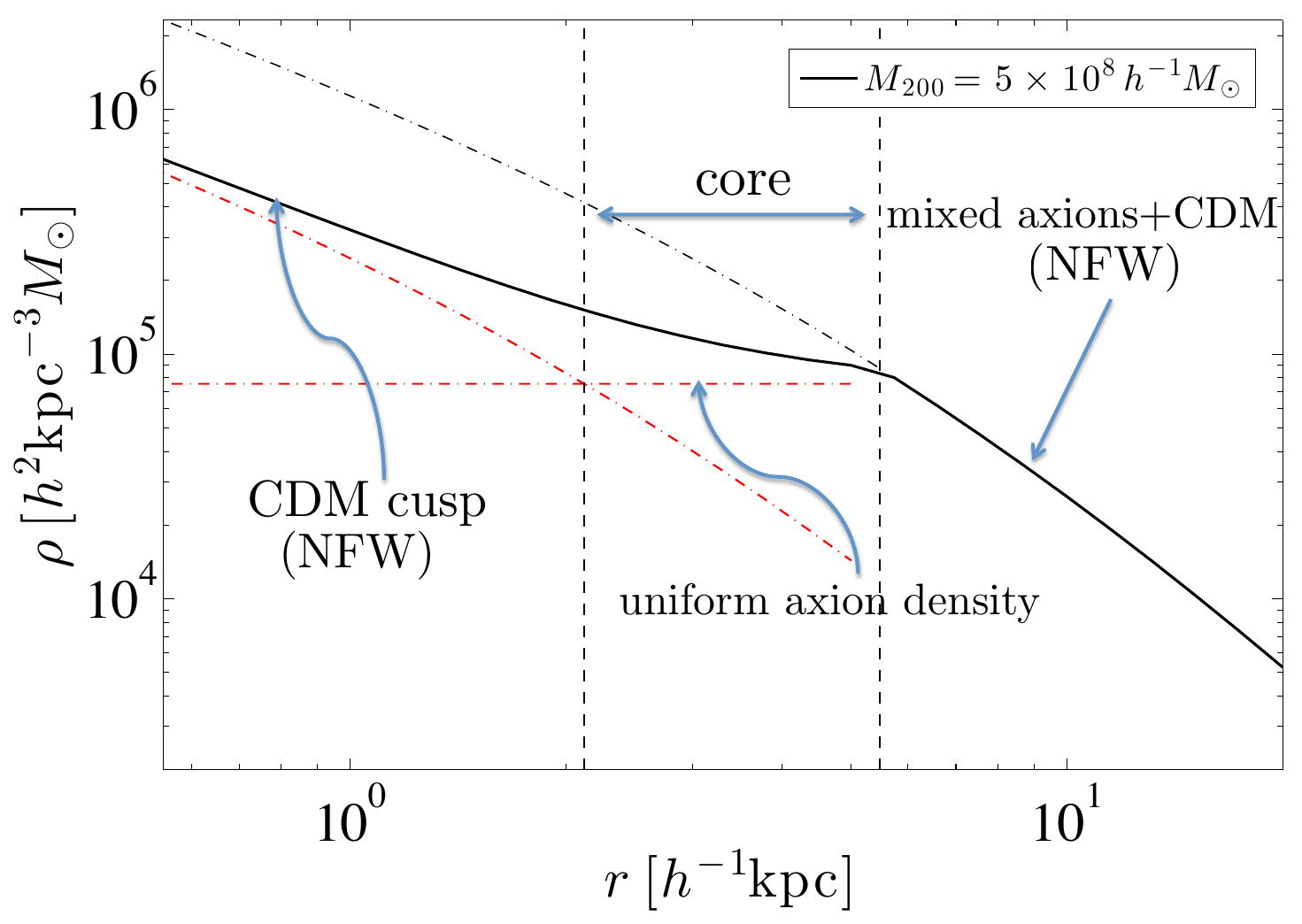} \\ [0.0cm]
 \end{array}$
 \caption{The mixed dark matter halo profile Eq.~(\ref{eqn:two_component_halo}) with $13\%$ CDM, $f_c=0.13$. The outer region is an NFW profile of mixed axions and CDM. The halo mass is $M_{200}=5\times 10^{8}h^{-1}M_\odot$ and the halo Jeans scale $r_{J,h}=5.4\,h^{-1}\text{ kpc}$ (outer vertical dashed line) corresponds to axion mass $m_a=10^{-22}\text{ eV}$ with this particular DM composition. At the halo Jeans scale the axions stop clustering and form a uniform component, while the CDM forms an NFW cusp. There is a core down to $r_{\rm core}=2.1\,h^{-1}\text{ kpc}$ where the cusp takes over from the uniform piece (inner vertical dashed line).}\label{fig:two_component_plot}
 % While this core is of significant size, the large value of $r_{\rm core}$ means such an ULA mass and aMDM composition may not provide an adequate resolution to the cusp-core problem of CDM.
\end{center}
 \end{figure}

\begin{figure}
$\begin{array}{@{\hspace{-0.1in}}l}
\includegraphics[scale=0.35]{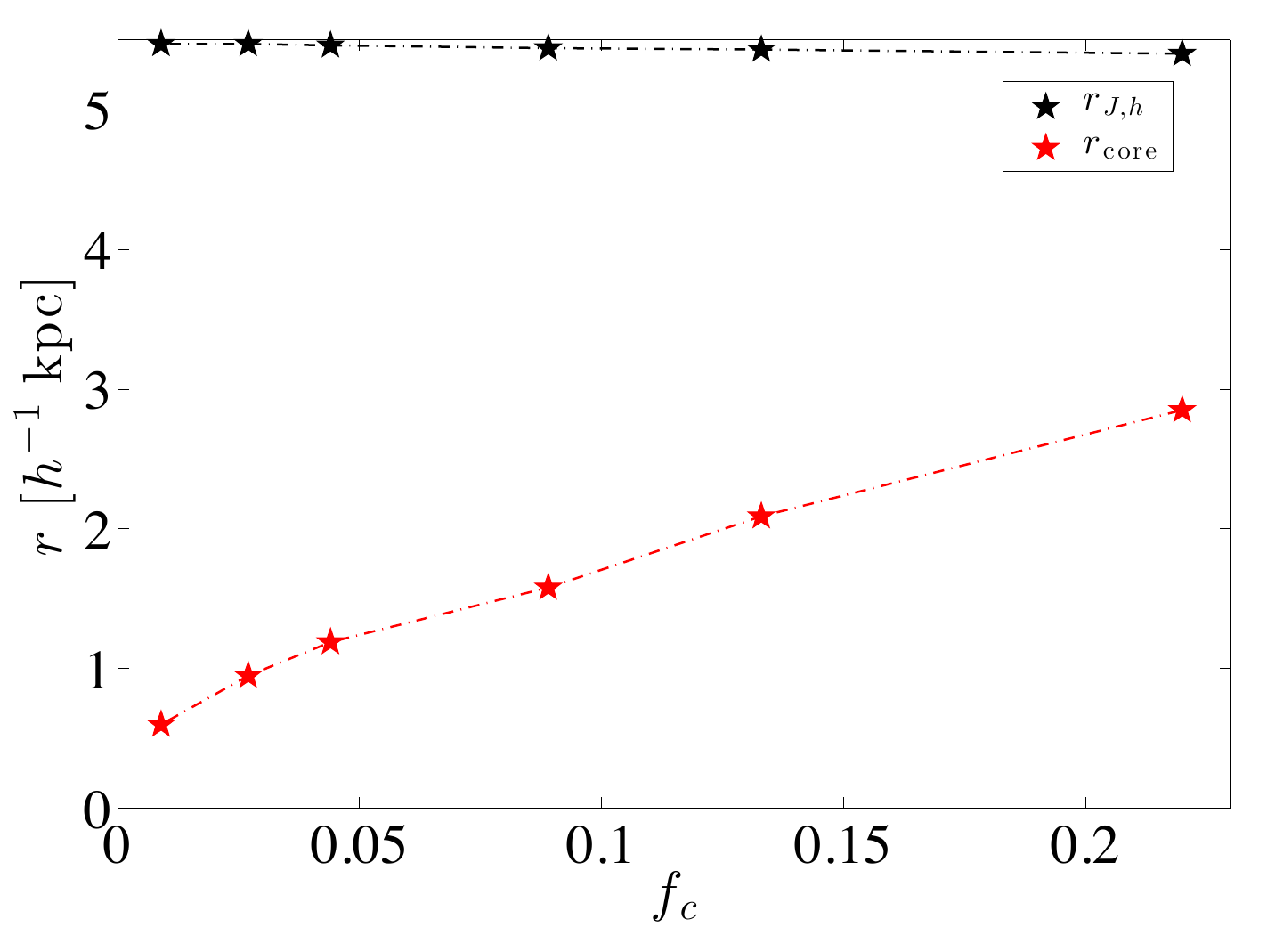} \\ [0.0cm]
 \end{array}$
\caption{Core radius and halo Jeans scale as a function of CDM fraction, $f_c=\Omega_c/\Omega_d$, in the two component halo (Eq.~(\ref{eqn:two_component_halo})). There is a core for $r_{\rm core}<r<r_{J,h}$, while there is a cusp for $r<r_{\rm core}$, so that increasing $f_c$ makes halo profiles more cuspy, as expected. The halo mass $M_{200}=5\times 10^{8}h^{-1}M_\odot$ and the axion mass $m_a=10^{-22}\text{ eV}$.}\label{fig:core_radius_fc}
\end{figure}

The size of the cored region depends on the fraction of DM that is cold: $f_c=\Omega_c/\Omega_d$. For $r<r_{\rm core}$ (see Eq.~(\ref{eqn:two_component_core_size})) there is a cusp while for $r_{\rm core}(f_c,M_s)<r<r_{J,h}(M_s)$ there is a core. We plot the halo profile for $f_c=0.13$ in our benchmark halo in Fig.~\ref{fig:two_component_plot}, while we plot $r_{\rm core}(f_c)$ for our benchmark halo in Fig.~\ref{fig:core_radius_fc}. We judge the core to be of significant size if it persists down to $<50\%$ of the halo Jeans scale. With $f_c=0.13$ the benchmark axion mass and halo corresponds to a core in the range $2.1\, h^{-1}\text{ kpc}\lesssim r \lesssim5.4 \, h^{-1}\text{ kpc}$. 

Although this benchmark core is significant, it is actually not present on sub-kiloparsec scales, which suggests that while introducing a fraction of CDM with an axion of mass $10^{-22}\text{ eV}$ may raise the HMF for low mass dwarf galaxies to acceptable levels, it may not provide a totally adequate solution to the CCP. As we will see below, a more preferable solution to all the problems outlined may be given instead by a higher axion mass.

\section{Too Big To Fail?}
\label{sec:tbtf}

The so-called `Too Big To Fail' problem (here `Massive Failures Problem', MFP) was introduced by \cite{boylan-kolchin2011}. In $\Lambda$CDM there are predicted to be massive subhaloes of the milky way of high concentration and circular velocity that cannot host bright satellites, and are not observed. One astrophysical solutions to this problem is feedback \citep{garrison-kimmel2013}. WDM \citep{lovell2012} and C+WDM \citep{maccio2012b,medvedev2013} are also known to help this problem, since the flattened variance leads to later formation times and lower concentration for these most massive subhaloes. 

\begin{figure}
$\begin{array}{@{\hspace{-0.2in}}l}
\includegraphics[scale=0.4]{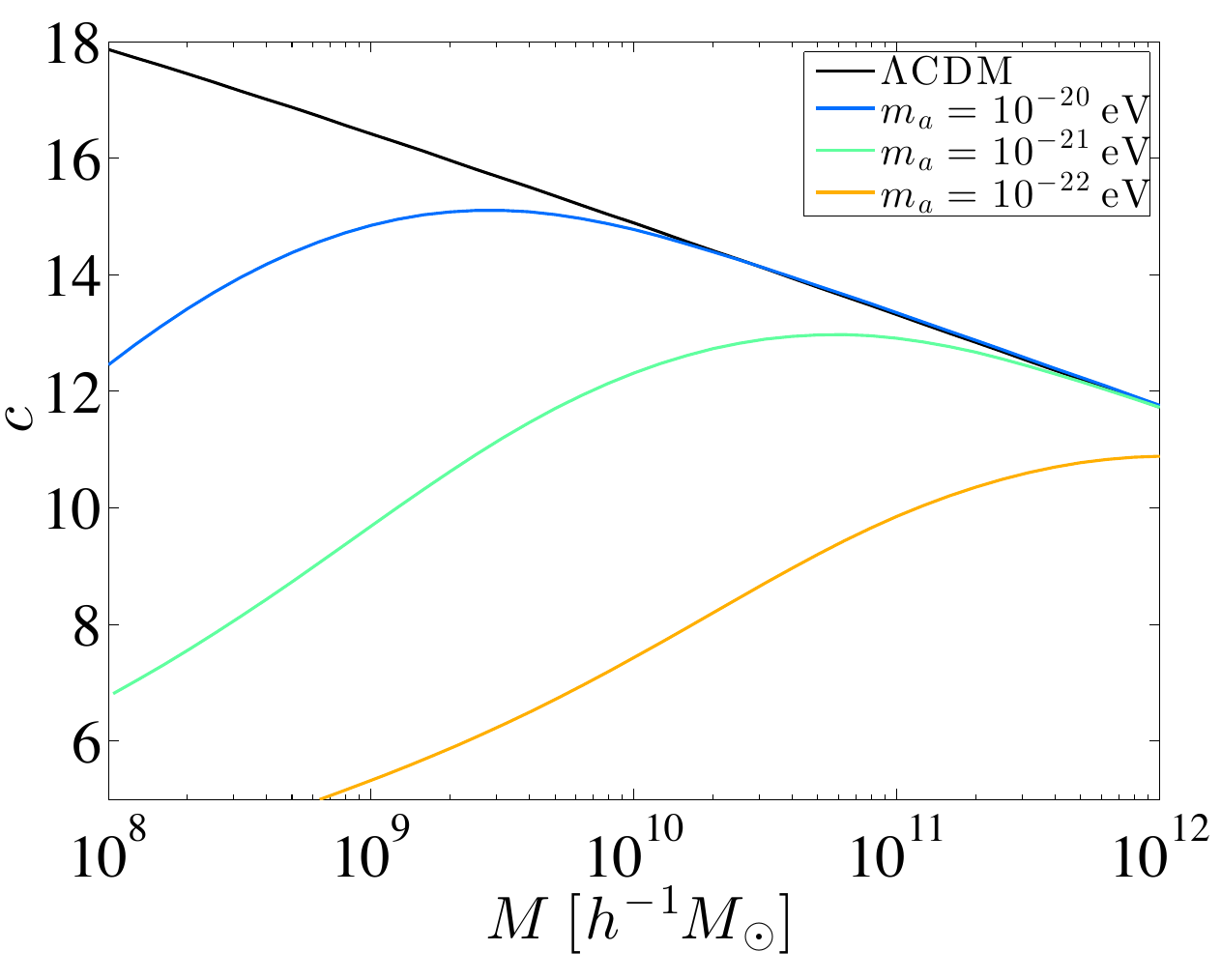} \\ [0.0cm]
 \end{array}$
\caption{Halo concentration parameter, $c(M)$ for $\Lambda$CDM and various axion masses. The flattening of the variance near $M_m$ causes a later formation time for low mass halos, and due to the presence of $f=0.01$ in Eq.~(\ref{eqn:zcoll_nfw}) leads to a lower concentration for halos below $f M_m$. The lowered concentration can help alleviate the `Too Big To Fail' problem.}\label{fig:con_m}
\end{figure}

Since the variance, $\sigma (M)$ in aMDM also flattens at low masses, just like WDM it will lead to a lower concentration for Milky Way sub-haloes compared to $\Lambda$CDM if the subhalo mass is lower than $f M_m$ (see Eq.~(\ref{eqn:zcoll_nfw})). In Fig.~\ref{fig:con_m} we plot $c(M_{200})$ for three representative axion masses, $m_a=10^{-22},10^{-21}10^{-20}\text{ eV}$ and compare to $\Lambda$CDM, confirming that this is the case.

\begin{figure}
\includegraphics[scale=0.36]{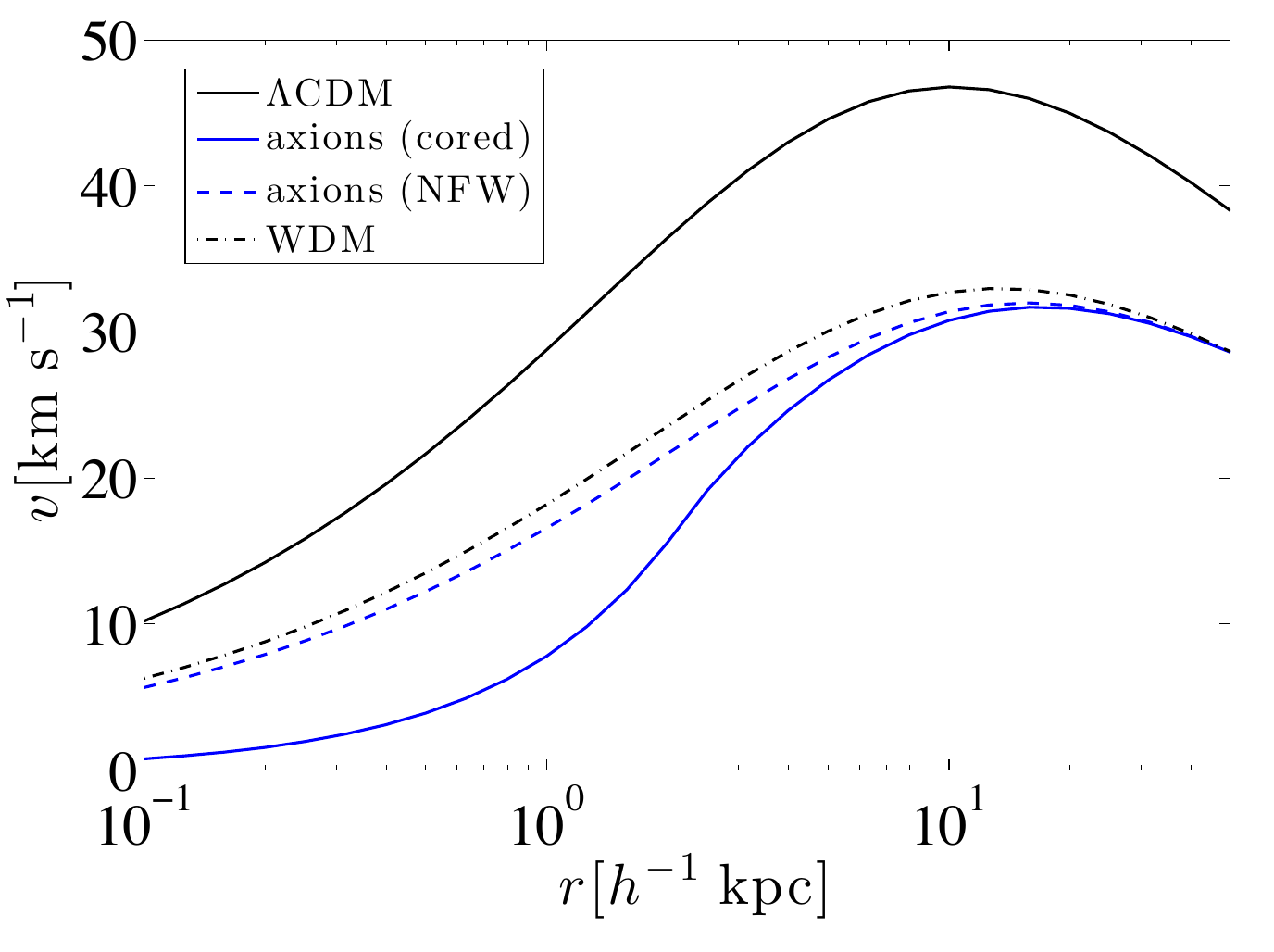}
\caption{The circular velocity profile for a one component DM halo in various models, derived from the NFW profile. For $\Lambda$CDM we have chosen a halo with $M_{200}=2\times 10^{10}\, h^{-1}M_\odot$ which has a maximum circular velocity $v_{\rm max}>40\text{ km s}^{-1}$, demonstrating that CDM suffers from the `Too Big To Fail' problem. We compare to axion and WDM models with $M_{200}=10^{10}\, h^{-1}M_\odot$ chosen such that they have the same $r_{\rm max}$ as $\Lambda$CDM. Both axions and WDM suppress $v_{\rm max}$ by a factor of about $1.5$ relative to $\Lambda$CDM, demonstrating their relevance for the solution to MFP. The core in the axion density profile does not affect the suppression of $v_{\rm max}$. We choose axion mass $m_a=10^{-22}\text{ eV}$ and equivalent WDM mass of $m_W\approx 0.84 \text{ keV}$.}\label{fig:vcirc}
\end{figure}

MFP can also be expressed as the non-observation of large numbers of satellites with maximum circular velocity $v_{\rm max}\gtrsim 40 \text{ km s}^{-1}$. The circular velocity as a function of $r$ is given by
\begin{equation}
v(r)=(G M(<r)/r)^{1/2} \, .
\end{equation}
In Fig.~\ref{fig:vcirc} we plot $v(r)$ for a halo of mass $M_{200}=2\times 10^{10}\, h^{-1}M_\odot$. In $\Lambda$CDM this halo has $v_{\rm max}>40 \text{ km s}^{-1}$. We compare this to $v(r)$ where the DM is made up of an axion of mass $m_a=10^{-22}\text{ eV}$, and to WDM. The axion and WDM halos are chosen to have the same $r_{\rm max}$ as $\Lambda$CDM, having $M_{200}=10^{10}\, h^{-1}M_\odot$. Both the axion and WDM models reduce the maximum circular velocity considerably: $v_{\rm max}\approx 30 \text{ km s}^{-1}$.

For WDM we do not model the effect of a possible core, and so for fair comparison we show the axion model with both a cored and NFW profile. The effect of the core in the axion model is small, so that the assumption of core formation does not affect an axion solution to MFP.

\section{High Redshift Objects}
\label{sec:collapse}

%\tkDM{Look at this with different fractions too? Perhaps just with the lightest axion to relieve tension...}
We now move on to discuss the collapse redshift of objects to see whether ULAs can accommodate observations of high-redshift galaxies. We consider collapse with $\Lambda$CDM growth given by $D(z)$, ignoring for the moment the scale-dependent growth, which will only serve to amplify effects below the characteristic mass. All the effects of axions therefore come in the variance, $\sigma(M)$. For the large axion fractions relevant for core formation in the two-component halo of Section~\ref{sec:two_component_profile} the effects of CDM on the variance are virtually negligible, so in our examples we take $\Omega_a/\Omega_d=1$ and investigate only the effects of varying axion mass.

\begin{figure}
$\begin{array}{@{\hspace{-0.1in}}l}
\includegraphics[scale=0.35]{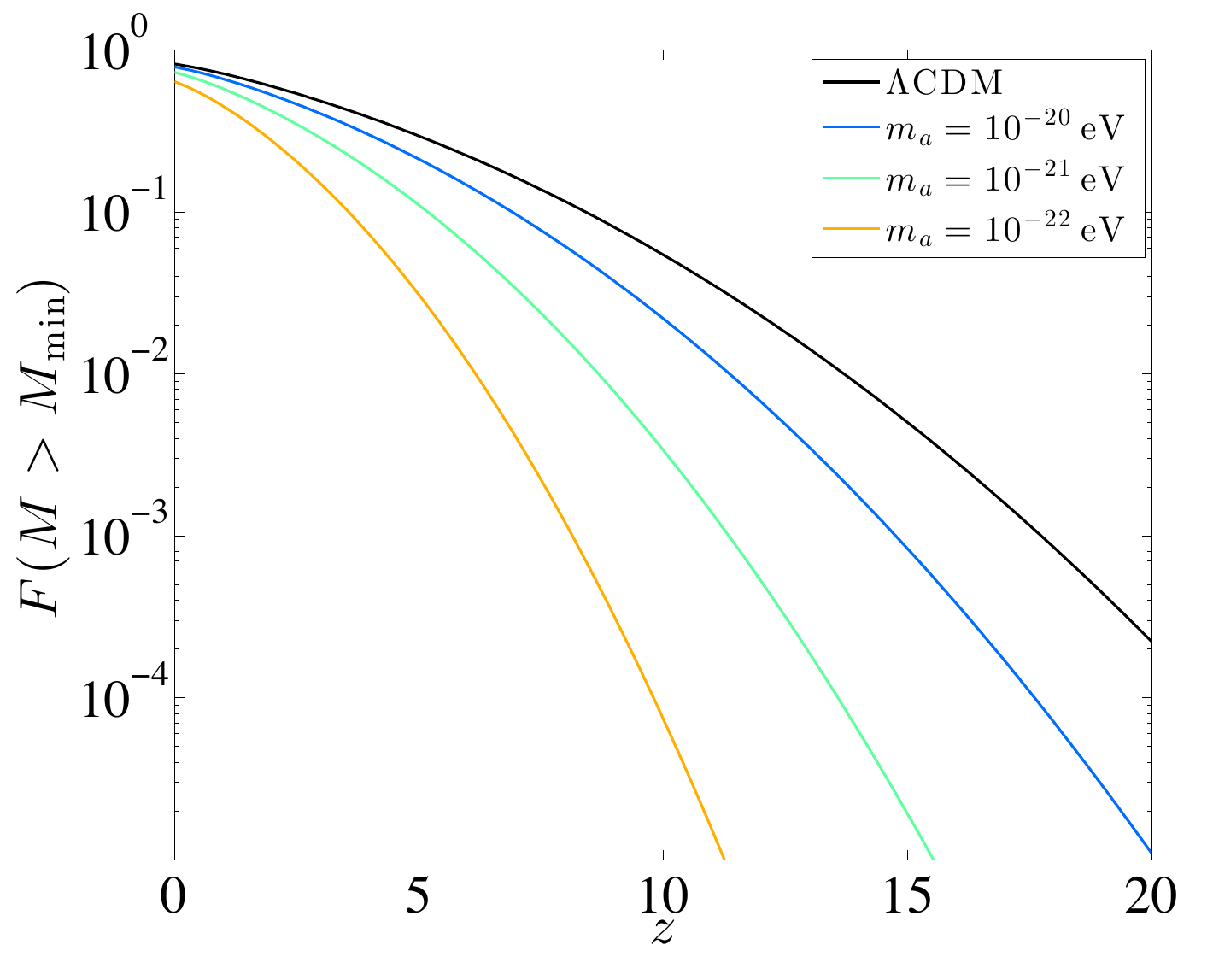} \\ [0.0cm]
 \end{array}$
\caption{Total collapsed mass fraction, Eq.~(\ref{eqn:coll_frac}), with $M_{\rm min}=10^6 \, h^{-1}M_\odot$ with varying axion mass. We ignore scale-dependent growth and take the axions to make up all of the DM.}\label{fig:coll_frac}
\end{figure}

The total fraction of objects collapsed with $M>M_{\rm min}$ at redshift $z$ is
\begin{equation}
F(M>M_{\rm min},z)=\text{erfc}\left(\frac{\delta_{c,\text{EdS}} D(z_{\text{coll}})^{-1}}{\sqrt{2}\sigma(M_{\rm min})} \right) \, .
\label{eqn:coll_frac}
\end{equation}
We plot this assuming $\Lambda$CDM growth for $M_{\rm min}=10^6 \, h^{-1}M_\odot$ in Fig.~\ref{fig:coll_frac}. For our benchmark mass $m_a=10^{-22}\text{ eV}$ the collapsed mass fraction $F(z\gtrsim 6)\lesssim 0.01$ putting such a light axion in considerable tension with observations of high redshift galaxies. 
%The filtering scale in the linear power spectrum causes the variance to flatten off below the characteristic mass, $M_m$, as was shown already in Fig.~\ref{fig:sigma_m_1e-22}, causing the argument of the error function in Eq.~(\ref{eqn:zcoll_nfw}) to go to zero and driving the redshift of collapse thus defined $z_{\rm coll}(M\ll M_m)\rightarrow 0$. Since we take $f=0.01$ in Eq.~(\ref{eqn:zcoll_nfw}), masses $M<100 M_m$ will be effected by the flattening off of the variance. This suitably reduces the concentration of low mass halos, but cannot really describe collapse in the axion model since structure formation described by barrier crossing will always be hierarchical with a monotonic variance \tkDM{as we demonstrate in NEW FIG}.

Using the scale-dependent growth to assign a mass-dependent critical density for collapse as in Eq.~(\ref{eqn:g_def}) provided a good working model for the cut-off in the HMF, however we cannot naively apply such a prescription for $\delta_c$ into Eq.~(\ref{eqn:coll_frac}): such a cut-off would make the collapsed mass fraction non-monotonic. The ansatz of Eq.~(\ref{eqn:g_def}) applies only in the HMF and takes the HMF as fundamental, so the analog of Eq.~(\ref{eqn:coll_frac}) is properly defined in this framework as the integral of the HMF. The redshift dependence of such an integral requires us to know the full function $D(k,z)$. Alternatively one could make a fit for the HMF with the ansatz that the cut-off remains always a fixed geometric distance between $M_m$ and $M_J$ as they evolve with redshift. We do not explore this effect of the cut-off on the collapsed mass fraction as a function of redshift, but the qualitative effects are obvious: scale-dependent growth should amplify the effects we have seen already in Fig.~\ref{fig:coll_frac}. 

Knowing the HMF at $z=10$ allowed the authors of \cite{pacucci2013} to place constraints on a WDM thermal relic of $m_W>0.9\text{ keV}$ using high redshift observations. Using our mass scale conversions of Section~\ref{sec:mass_scales} we might expect such observations to constrain $m_a\gtrsim \text{few}\times 10^{-22}\text{eV}$, as the simple argument based on collapsed mass fraction with $\Lambda$CDM growth given above anticipates.

Suppression of galaxy formation at high redshift has been invoked as a possible solution to another problem of structure formation in $\Lambda$CDM: the discrepancy in the evolution of the stellar mass function between observations and models, highlighted in \cite{weinmann2012}. ULAs were invoked, along with WDM, by these authors as a solution. Again, as with Lyman-$\alpha$ constraints, access to hydrodynamical simulations with WDM allowed for a detailed comparison of models to observations, and showed a WDM solution to most likely be unviable. Do ULAs remain a viable solution? We have shown in previous sections that there are enough differences between ULAs and WDM that this is possible, but without simulations one cannot quantify this. However, in this section we have confirmed the suppression of halo formation at high redshift necessary for ULAs to be an interesting candidate for further study in this regard.

\section{Summary and Discussion}
\label{sec:discussion}

By studying large scale structure we have probed ultra-light axion-like-particles (ULAs) with masses in the range $10^{-24}\text{ eV}\leq m_a\leq 10^{-20}\text{ eV}$. Across a fair portion of this range such ultra-light fields can evade large-scale structure constraints while still being different enough from standard CDM on scales relevant to three main problems of structure formation: the missing satellites problem (MSP), the cusp-core problem (CCP), and the massive failures problem (MFP). We have primarily studied a benchmark ULA  of mass $m_a=10^{-22}\text{ eV}$ and shown that it is able to solve the MSP, CCP and MFP, avoiding the so-called \emph{Catch 22} of a WDM solution. If this axion constitutes all of the DM, however, then it may come into tension with observations of the Lyman-$\alpha$ flux power spectrum (which constrains WDM at masses $m_W\gtrsim 3.3\text{ keV}$), high-redshift galaxies (at $z\gtrsim 6$) and the existence of very low mass dwarf galaxies ($M\lesssim 5\times 10^7\, h^{-1}M_\odot$). These tensions can be relieved in two ways: by introducing a fraction of CDM or increasing the axion mass. Introducing a fraction of CDM retains adequate solutions to all problems, but cores may yield to cusps at unacceptably large radii. We advocate a higher mass axion of $m_a\gtrsim 10^{-21}\text{ eV}$ as potentially the best solution.

If all the DM is constituted of an axion or other light scalar of mass $m_a=10^{-22}\text{ eV}$ then in the linear power spectrum structure formation is suppressed below some characteristic scale (Fig.~\ref{fig:lin_tk}). The half-mode for this suppression is the same as the half-mode for a WDM particle of mass $m_W\approx 0.84 \text{ keV}$, just on the edge of the bounds coming from Lyman-$\alpha$ forest flux power spectrum constraints (Fig.~\ref{fig:mw_ma}). By considering scale-dependent growth it has been shown that such an axion will cut off the halo mass function for $M\lesssim 10^8\, h^{-1}M_\odot$. Introducing a fraction of CDM, $f_c=\Omega_c/\Omega_d$, the cut-off is made less severe, disappearing completely and leaving only a small suppression to dwarf galaxy formation when $f_c\approx 0.5$ (Fig.~\ref{fig:hmf_deltac_m}). This mixed dark matter model, aMDM, may therefore be relevant to the MSP. Such a mix of DM may be natural given certain theoretical priors \citep{aguirre2004,wilczek2004,tegmark2006,bousso2013,wilczek2013}, or in non-thermal cosmologies expected after moduli stabilisation \citep[e.g.][]{acharya2010a,acharya2010b}.

If the axion linear Jeans scale can be considered to scale into non-linear environments as the fourth root of the relative density contrast (Eqs.~(\ref{eqn:hu_jeans_def}), (\ref{eqn:jeans_halo})), requiring coherence of field oscillations to be maintained, then ultra-light axions can give cores to dwarf galaxy density profiles \citep{hu2000}. By considering a simple cored profile given by an external NFW profile outside the Jeans scale, it was shown that in such a scenario no halos are formed below the linear Jeans mass, $M_J$. Axions in the mass range $10^{-22}\text{ eV}\leq m_a\leq 10^{-20}\text{ eV}$ can give kiloparsec scale cores to dwarf galaxies of mass $M=5\times 10^{8}\, h^{-1}M_\odot$, and are thus relevant to the CCP of CDM halo density profiles (Figs.~\ref{fig:one_component_core_profile}, \ref{fig:rjh_axionmass}). Since the halo mass function is only cut off below the dwarf mass, and axions in this mass range are allowed by large-scale structure constraints, we can conclude that in this simple model ultra-light axions, or other `Fuzzy' CDM candidates, do not suffer from the \emph{Catch 22} that might affect WDM \citep{maccio2012}.

For an axion mass at the low end of the range allowed by large-scale structure constraints, $m_a\approx 10^{-22}\text{ eV}$, to form lighter dwarf galaxies of $M\lesssim 5\times 10^{7}\, h^{-1}M_\odot$, however, the mixed aMDM is necessary. By considering a two-component density profile below the axion Jeans scale it was shown that an admixture of $f_c\approx 13\%$ will significantly increase the mass function for light dwarves (Fig.~\ref{fig:hmf_cutoff_1e-22}), while still allowing for a core on scales greater than a kiloparsec (Fig.~\ref{fig:two_component_plot},~\ref{fig:core_radius_fc}), although such a core may in fact be too large.

The flattened variance in aMDM (Fig.~\ref{fig:sigma_m_1e-22}) in the NFW formalism leads to later formation times and consequently lower concentrations for low mass halos compared to CDM (Fig.~\ref{fig:con_m}). This, combined with maximum circular velocity remaining low, $v_{\rm max}<40\text{km s}^{-1}$, in typical dwarves (Fig.~\ref{fig:vcirc}), also suggests that ULAs may, just like WDM, play a role in the resolution of the MFP \citep{boylan-kolchin2011}. 

While avoiding the \emph{Catch 22} in an axion cusp-core and missing satellites resolution an axion mass as low as $m_a=10^{-22}\text{ eV}$ comes into tension with high-redshift observations since the collapsed mass fraction becomes very small at $z\gtrsim 6$ (Fig.~\ref{fig:coll_frac}). A heavier axion of mass $m_a \gtrsim 10^{-21}\text{ eV}$ would be in less tension with observations of high-redshift galaxies (and more recent Lyman-$\alpha$ forest constraints to WDM) and could still introduce a kiloparsec scale core to dwarf galaxies and significantly lower the concentration of these galaxies. The formation of dwarves would still be significant, yet also reduced relative to $\Lambda$CDM, so that such heavier axions remain relevant to the cusp-core, missing satellites, and `too big to fail' problems of CDM. Suppression of galaxy formation at high redshift relative to $\Lambda$CDM may also be a factor in resolving conflicts between models and observations of the stellar mass function \citep{weinmann2012}.

With large axion fractions \cite{marsh2013} showed that isocurvature constraints imply such a model would be falsified by any detection of tensor modes at the percent level in the CMB by Planck. Axions lighter than those we study are well constrained as components of the DM by observations of the CMB and the linear matter power spectrum \citep{amendola2005,marsh_inprep}, while heavier axions are probed, and in some cases ruled, out by the spins of supermassive black holes \citep{arvanitaki2010,pani2012}, and terrestrial axion searches \citep{jaeckel2010b,ringwald2012}.

We have not discussed the role of baryons in this model, the knowledge of this role being incomplete in even the most state of the art simulations. We have preferred to focus on simple and idealised DM only models where the relevant physics is well understood. The baryonic disk has only a modest effect on the rotation curve, with DM halos still necessary. Adiabatic contraction of baryons may lead to the enhancement of DM cusps, and thus more need for a core forming component \cite{zemp2012}. On the other hand, baryonic feedback is a process driven by supernova explosions driving outflows of gas, which can remove DM cusps in dwarf galaxies while leaving a thick stellar disk \cite{governato2012}. Baryons may also transfer angular momentum to the halo and modestly effect the spin-up of a massive halo. In massive galaxies it is unlikely that baryons have a significant effect on the DM halo profile. None of these effects, most notably feedback, solve the excess baryon problems of the MSP and MFP.

It will be important to investigate this model further in the future with numerical N-body and other non-linear studies in order to verify whether our simple predictions stand up to detailed scrutiny both theoretically and observationally. It is possible that non-linear effects such as oscillons (see e.g. \cite{gleiser2009}) may play a role. Also, at non-linear order additional terms in the effective fluid description of the axion will be generated, such as anisotropic stresses \cite{hertzberg2012}, which could alter the simple picture of structure formation with a sound speed and Jeans scale dominating effects at short distances.

Furthermore, Lyman-$\alpha$ constraints play a key role in determining the validity of WDM models to resolve small scale crises in CDM, the constraints of \cite{viel2013} appearing to all but rule out WDM in this regard. Applying such constraints reliably to ULAs will require developing hydrodynamical simulations with them. Finally, a thorough development of the halo model with ULAs building on the groundwork laid here (as was done for WDM by \cite{smith2011}) will be invaluable in understanding weak lensing constraints to ULAs obtainable with future surveys \citep{marsh2011b,euclid_theory_report}.

% ------------------------ ACKNOWLEDGEMENTS ----------------------------------
\section*{Acknowledgments}

DJEM acknowledges the hospitality of the BIPAC at Oxford University, where this work was initiated, and of Institute d'Astrophysique de Paris, where it was completed. DJEM also acknowledges useful discussions with Andrew Pontzen, Daniel Grin, Scott Tremaine, Enrico Pajer, Jesus Zavala Franco, Simeon Bird and Tom Abel, and an email exchange with Robert Smith. We thank Ren\'{e}e Hlozek for comments on the manuscript, and the anonymous referee for helpful comments and reference suggestions. Research at Perimeter Institute is supported by the Government of Canada through Industry Canada and by the Province of Ontario through the Ministry of Research and Innovation.  The research of JS has been supported at IAP by  the ERC project  267117 (DARK) hosted by Universit\'e Pierre et Marie Curie - Paris 6   and at JHU by NSF grant OIA-1124403.

 \renewcommand{\theequation}{A\arabic{equation}}
  % redefine the command that creates the equation no.
  \setcounter{equation}{0} 
\appendix
\section{Details of the Mixed Dark Matter Profile}
\label{appendix:normalisation}

In Section~\ref{sec:jeans_and_norm} we discussed the simple cored profile. The two component profile of Section~\ref{sec:two_component_profile} is given by
\begin{align}
\rho_{\rm aMDM}(M,&r,f_c)= \theta (r-r_{J,h}(M_s))\rho_{\rm NFW}(M_s,r)\nonumber\\
	+&\theta (r_{J,h}(M_s)-r) [(1-f_c)\rho_{\rm NFW} (M_s, r_{J,h}(M_s))\nonumber \\
	+&\rho_{\rm NFW} (M_{\star}(M_s,f_c),r)] \, , 
\label{eqn:two_component_halo}
\end{align} 
where $\theta(x)$ is the Heaviside function and $f_c=\Omega_c/\Omega_d$. $M_s\neq M_{200}$ is the mass of the external NFW profile, the `scale mass'. The scale mass of the internal NFW profile for the CDM, $M_{\star}(M_s,f_c)$, is fixed by continuity and the requirement that at the halo Jeans scale the ULAs and CDM are in their relative cosmic abundance
\begin{equation}
\rho_{\rm NFW}(M_{\star}(M_s,f_c),r_{J,h}(M_s))=f_c \rho_{\rm NFW}(M_s,r_{J,h}(M_s)) \, .
\end{equation}
The CDM NFW profile inside the axion Jeans scale must be assigned a concentration, which must be computed from the variance in a given cosmology. We take the appropriate variance to be the one for the cosmology as a whole. This assumes that, with hierarchical structure formation, this CDM inner region will be made from a lighter halo itself formed earlier in cosmic history. 

To find the size of the cored region for a given CDM fraction $f_c$ we simply find the radius $r_{\rm core}$ that solves
\begin{equation}
(1-f_c)\rho_{\rm NFW} (M_s,r_{J,h}(M_s))=\rho_{\rm NFW} (M_\star(M_s,f_c),r_{\rm core}) \, .
\label{eqn:two_component_core_size}
\end{equation}
As $f_c\rightarrow 1$, $r_{\rm core}\rightarrow r_{J,h}(M_s)$, the cored region disappears and Eq.~(\ref{eqn:two_component_halo}) goes to the standard NFW case. 

%\section{CMB Spectral Distortions}
%\label{appendix:spectral_distortions}

%ULAs have a model-dependent coupling to $\vec{E}\cdot \vec{B}$. If there are primordial magnetic fields, the vacuum realigment production of axions at relatively late times could produce photons, leading to spectral distortions of the CMB. In this section we investigate this by coupling the Klein-Gordon equation (\ref{eqn:klein_gordon}) to Maxwell's equations in a radiation dominated universe.

 % ---------------------- BIBLIOGRAPHY -----------------------------------------

\bibliography{doddyoxford,silk_bibliography}
\bibliographystyle{mn2e.bst}

\end{document}